\documentclass{ws-ijmpd}
\usepackage{amssymb}
\usepackage{graphicx}
\usepackage{rotating}
\begin{document}
%

\markboth{R.~Aurich et al.}
{Ellipticity of Structures in CMB Sky Maps}

\catchline{}{}{}{}{}

\title{Ellipticity of Structures in CMB Sky Maps}

\author{Ralf Aurich, Holger S.\ Janzer, Sven Lustig}

\address{Institut f\"ur Theoretische Physik, Universit\"at Ulm,\\
Albert-Einstein-Allee 11, D-89069 Ulm, Germany}

\author{Frank Steiner}

\address{Universit\'e Lyon 1,
Centre de Recherche Astrophysique de Lyon\\ CNRS UMR 5574,
9 avenue Charles Andr\'e, F-69230 Saint-Genis-Laval, France}

\maketitle

\begin{abstract}
We study the ellipticity of contour lines in the sky maps
of the cosmic microwave background (CMB)
as well as other measures of elongation.
The sensitivity of the elongation on the resolution of the CMB maps
which depends on the pixelization and the beam profile of the detector,
is investigated.
It is shown that the current experimental accuracy does not allow to
discriminate between cosmological models which differ in curvature by
$\Delta \Omega_{\hbox{\scriptsize tot}}=0.05$.
Analytical expressions are given for the case
that the statistical properties of the CMB are those
of two-dimensional Gaussian random fields.
\end{abstract}




\section{Introduction}


The cosmic microwave background (CMB)  gives a wealth of information
about the Universe.
The usual statistical analysis comprises the angular auto-correlation function
$C(\vartheta)$ as well as the angular power spectrum $\delta T_l^2$,
which lead to an improved determination of the cosmological parameters.
Besides these statistics there are topological/geometrical descriptors
of the CMB
such as the number densities of maxima and minima,
the ellipticities of the peaks, and peak correlation properties.
Under the assumption that the CMB has the statistical properties
of two-dimensional homogeneous and isotropic Gaussian random fields,
the theoretical predictions are derived in Ref.~\refcite{Bond_Efstathiou_1987}.

This paper deals with the structure of the contour lines of a given
temperature level $\delta T$ in the CMB maps.
In the vicinity of extrema these contour lines can be approximated by
ellipses having the ellipticity $E$
\begin{equation}
\label{Eq:Ellipticity}
E \; := \; \frac{a}{b}
\hspace{10pt} , \hspace{10pt}
E \; \ge \; 1
\hspace{10pt} , 
\end{equation}
where $a \ge b$ denote the major and minor semi-axes of the ellipses
which best match the contour line around a given peak.
The interesting question is how the ellipticities are distributed for a
given map in dependence on the temperature level $\delta T$.
This provides a further test\cite{Dore_Colombi_Bouchet_2003,%
Monteserin_Barreiro_Martinez-Gonzalez_Sanz_2006}
of Gaussianity by comparing the ellipticities
with the corresponding results of two-dimensional Gaussian random
fields\cite{Bond_Efstathiou_1987}.
In Refs.~\refcite{Barreiro_Sanz_Martinez-Gonzalez_Cayon_Silk_1997,%
Barreiro_Martinez-Gonzalez_Sanz_2001,%
Monteserin_Barreiro_Sanz_Martinez-Gonzalez_2005}
it is shown, however, that there are better geometrical estimators
as tests of Gaussianity
than the ellipticity like the Gaussian curvature.
Therefore, the main motivation for this paper is to study the discrepancy
of the ellipticity between the observations and simulations of the CMB
as reported in
Refs.~\refcite{Gurzadyan_Torres_1997,Gurzadyan_et_al_2003a,Gurzadyan_et_al_2003b,%
Gurzadyan_et_al_2004,Gurzadyan_et_al_2005,Gurzadya_et_al_2007}.
In these papers an excess in the ellipticity $E=2.2\dots 2.7$
is found compared to standard
$\Lambda$CDM models having $E$ around 1.65.
Due to noise and the foreground uncertainties
it is difficult to obtain the ellipticity from the measured CMB radiation.
The ellipticity was determined from the
COBE-DMR map\cite{Gurzadyan_Torres_1997}
and from the BOOMERanG map\cite{Gurzadyan_et_al_2003a,Gurzadyan_et_al_2003b}.
These results are confirmed by the analysis of the
WMAP 1yr data\cite{Gurzadyan_et_al_2004,Gurzadyan_et_al_2005}
and the 3yr data\cite{Gurzadya_et_al_2007}.
Furthermore, a strong dependence of the ellipticity on the curvature
of the Universe is claimed and, with respect to the excess ellipticity,
this is interpreted as a hint towards a hyperbolic Universe.
In Ref.~\refcite{Barreiro_Sanz_Martinez-Gonzalez_Cayon_Silk_1997}
there is no such ellipticity-curvature correlation found by studying
low $\Omega$ and flat universes based on CDM models.
Thus there remains the question whether the excess ellipticity is real
and how such an excess, if present, has to be interpreted.


\section{CMB Sky Maps}


\subsection{Cosmological models}
\label{Cosmological model}

The standard $\Lambda$CDM model of cosmology is given by
a certain set of cosmological parameters.
The curvature of the universe is determined by the value of the
total energy density $\Omega_{\hbox{\scriptsize tot}}$ 
at the present epoch which is the sum
\begin{equation}
\label{Eq:Omega_0}
\Omega_{\hbox{\scriptsize tot}} \; = \;
\Omega_{\Lambda} + \Omega_{\hbox{\scriptsize cdm}} +
\Omega_{\hbox{\scriptsize b}} + \Omega_{\hbox{\scriptsize r}}
\end{equation} 
of the energy density $\Omega_{\Lambda}$ of dark energy,
$\Omega_{\hbox{\scriptsize cdm}}$ of cold dark matter,
$\Omega_{\hbox{\scriptsize b}}$ of baryonic matter and
$\Omega_{\hbox{\scriptsize r}}$ of radiation.
A value $\Omega_{\hbox{\scriptsize tot}}>1$, $=1$ or $<1$
reveals a spherical, flat or hyperbolic universe,
respectively.
The concordance $\Lambda$CDM model\cite{Spergel_et_al_2006,Komatsu_et_al_2003}
leads to a flat universe
where the angular power spectrum $\delta T_l^2 = l(l+1) C_l/(2\pi)$
has the first acoustic peak at $l\simeq 220$.
The following analysis investigates the dependence of the ellipticity
on the curvature of the universe.
To that aim only the amount of dark energy $\Omega_{\Lambda}$ is varied,
all other cosmological parameters are those of the
$\Lambda$CDM concordance model.

In addition to the infinite volume $\Lambda$CDM concordance model,
a model with a cubic topology\cite{Aurich_Janzer_Lustig_Steiner_2007,%
Aurich_2008}
is also studied which has a finite volume and is statistically anisotropic.
This model thus violates some conditions required
for the following analytical expressions as outlined in the Appendix.
This model serves as a test whether an anisotropic non-trivial topology can
be discerned by an elongation measure.
The sky maps are simulated for the cubic topology using the
same cosmological parameters as for the infinite concordance model.
The side length $L$ of this toroidal topology is chosen as $L=3.86\,L_H$
where $L_H$ is the Hubble length.
For this length $L$ a better agreement with the correlations
observed in the CMB sky is found\cite{Aurich_Janzer_Lustig_Steiner_2007,%
Aurich_2008,Aurich_Lustig_Steiner_2009}
than for the $\Lambda$CDM concordance model.
The eigenmodes belonging to the first 50\,000 eigenvalues for
the cubic topology are used for the simulation,
i.\,e.\ a total of 61\,556\,892 eigenmodes.
Furthermore, the eigenmodes are expanded in the
spherical basis up to $l_{\hbox{\scriptsize max}}=1000$
yielding sky maps with structures beyond the third acoustic peak
where the Silk damping already smoothes a further fine-structure.
In this way CMB sky simulations are obtained
for which resolution effects can be studied.

For a given cosmological model, i.\,e.\ a given set of cosmological parameters,
one can calculate the multipole spectrum
\begin{equation}
\label{Eq:powerspectrum}
C_l \; := \;
\left\langle \frac{1}{2l+1} \sum\limits_{m=-l}^{l}
\left|a_{lm}\right|^2\right\rangle
\hspace{10pt} ,
\end{equation} 
where $a_{lm}$ are complex coefficients obtained from the expansion
of a CMB sky map $\delta T(\theta,\phi)$ into
spherical harmonics $Y_{lm}(\phi,\theta)$ due to 
\begin{equation}
\label{Eq:expansion}
\delta T(\theta,\phi) \; = \;
\sum\limits_{l=0}^{\infty} \sum\limits_{m=-l}^{l} a_{lm} Y_{lm}(\phi,\theta)
\hspace{10pt} .
\end{equation}
The multipole spectrum $C_l$ is obtained from an ensemble average
denoted by $\left\langle \ldots\right\rangle$
over infinitely many realisations of universes with fixed
cosmological parameters. 
The deviation of the multipole spectrum $C_l$ of an individual
realisation from the ensemble average is characterised by the cosmic variance 
\begin{equation}
\label{Eq:var_powerspectrum}
{\hbox{Var}}(C_l) \; := \; \frac{2 C_l^2}{2 l+1} 
\hspace{10pt} ,
\end{equation}
where one has to assume
that the CMB is a homogeneous isotropic Gaussian random field.
This has to be taken into account
by comparing theory and experiment.

\subsection{Resolution of the sky maps}

Statistical measures of the niveau lines depend sensitively on the
resolution of the sky maps.
Thus a few remarks are in order.
A natural cut-off in the multipole space is provided by the physics of the CMB,
especially by the Silk damping and
the smoothing due to the thickness of the surface of last scattering.
One would need sky maps with a resolution of at least this physical cut-off
in order to capture all genuine CMB structures.
This is currently beyond the possibilities, and one is forced to consider
sky maps which are limited by the measurements,
i.\,e.\ by the beam profile of the detector.
The comparison of the measured sky map with a simulated map requires
that the simulation is accordingly smoothed.
A symmetric Gaussian smoothing kernel is sufficient for most applications
which is a special case of a general symmetric smoothing kernel
represented by the window function $F_l$.
The smoothing operation is done in multipole space by
\begin{equation}
\label{Eq:smooth1}
\delta T(\theta,\phi) \; \rightarrow \; \delta T_{F_l}(\theta,\phi) \; = \;
\sum_{l=0}^{\infty} \, \sum_{m=-l}^{l} F_l \, a_{lm} \, Y_{lm}(\theta,\phi)
\hspace{10pt} .
\end{equation}
The symmetric Gaussian kernel is given by the window function 
\begin{equation}
\label{Eq:gauss_kernel}
F_l^{\hbox{\scriptsize Gauss}} \; = \;
\exp \left[ - \frac{\sigma_{\hbox{\scriptsize g}}^2 \, l (l+1)}{2}\right]
\hspace{10pt} ,
\end{equation} 
where $\sigma_{\hbox{\scriptsize g}}$ is the 
width of the symmetric Gaussian kernel
which is usually parameterised by its full width at half maximum
$\sigma_{\hbox{\scriptsize fwhm}}$.
The conversion formula is given by
\begin{equation}
\label{Eq:fwhm}
\sigma_{\hbox{\scriptsize g}} \; = \;
\frac{\pi}{180^{\circ}} \; 
\frac{ \sigma_{\hbox{\scriptsize fwhm}}}{2\sqrt{2\ln (2)}}
\hspace{10pt} ,
\end{equation}     
where $\sigma_{\hbox{\scriptsize fwhm}}$ is given in degrees.

An analysis of the temperature fluctuation field in position space
requires a pixelization of the data on the sphere.
The CMB sky maps are usually discretized in the
HEALPix\cite{Gorski_Hivon_Banday_Wandelt_Hansen_Reinecke_Bartelmann_2005}
format where every pixel covers an equal area.
The resolution parameter $N_{\hbox{\scriptsize side}}$ defines the total number
of pixels by $N_{\hbox{\scriptsize pix}}^{\hbox{\scriptsize tot}} = 
12 \, N_{\hbox{\scriptsize side}}^2$.
The resolution in position space can be defined by the square root
of the area of a single pixel. 
The area of a pixel is
$10800/(\pi\,N_{\hbox{\scriptsize side}}^2)\hbox{ deg}^2$.

The WMAP data are available in $N_{\hbox{\scriptsize side}}=512$
at the LAMBDA website (lambda.gsfc.nasa.gov),
and the Planck data will be provided in $N_{\hbox{\scriptsize side}}=2048$.
As a rule of thumb the pixel resolution should relate to
the resolution in multipole space by 
$l_{\hbox{\scriptsize max}} = 2 N_{\hbox{\scriptsize side}}$.

\subsection{Contamination of CMB measurements}

A measurement gives the genuine CMB signal superimposed with
the emission of foreground sources and the detector noise.

Noise is unavoidable in measurements of CMB sky maps and acts mostly
on the smallest scales, i.\,e.\ at large multipoles $l$.
It can be modulated by generating in each pixel random fluctuations
which are added to the simulated CMB signal.
In the case of the WMAP data
the standard deviation of the random fluctuations is proportional to
$1/\sqrt{N_{\hbox{\scriptsize obs}}}$
where the constant of proportionality is stated on the LAMBDA website
and $N_{\hbox{\scriptsize obs}}$ is the number of observations 
of a given pixel\cite{Bennett_et_al_2003}.
Since $N_{\hbox{\scriptsize obs}}$ depends on the pixel,
the noise also depends on the direction yielding anisotropic noise properties.

Astrophysical radiation sources bring foreground contaminations
on measured CMB sky maps.
Cleaning operations reduce their contributions,
but cannot avoid that still some sky regions have to be excluded
from the analysis by masking them out.
This is a serious problem for estimators
which are based on full sky information as the angular power spectrum
$\delta T_l^2$.
For the structure analysis the remaining foreground contamination
in regions, that are not masked out,
is more important since it changes the properties of the niveau lines.


\section{Measures of Elongation}


Probably the most common measure of the elongation of a structure
is the ellipticity $E$
which is based on the assumption that contour lines are
approximated by ellipses.
$E$ is determined by the semi-axes $a$ and $b$,
which are real numbers with $a\geq b$,
of the best-fit ellipsis.
To parameterise the elongation,
various dimensionless combinations of the semi-axes can be used.
Three quantities are found in the literature and given by
the ellipticity $E$, the eccentricity $\varepsilon$,
and a further elongation measure $e$.
Their definitions and conversion formulas are given by
\begin{equation}
\label{Eq:def_ellip}
E \; := \; \frac{a}{b} \; = \; \sqrt{\frac{1 + 2e}{1 - 2e}} \; = \;
\frac{1}{\sqrt{1 - {\varepsilon}^2}}
\hspace{10pt} , \hspace{10pt}
1\le \; E \; < \; \infty
\hspace{10pt} ,
\end{equation}
\begin{equation}
\label{Eq:def_excen}
\varepsilon \; := \; \frac{\sqrt{a^2-b^2}}{a} \; = \;
\sqrt{1-\frac{1}{E^2}} \; = \; 2 \sqrt{\frac{e}{1 + 2e}}
\hspace{10pt} , \hspace{10pt}
0\le \; \varepsilon \; < \; 1
\hspace{10pt} \hbox{, \; and}
\end{equation}
\begin{equation}
\label{Eq:def_e}
e \; := \; \frac{1}{2} \, \frac{1-{\frac{b^2}{a^2}}} {1 + {\frac{b^2}{a^2}}}
\; = \; \frac{1}{2} \, \frac{{\varepsilon}^2}{2 - {\varepsilon}^2}
\; = \; \frac{1}{2} \, \frac{1-\frac{1}{E^2}}{1 + \frac{1}{E^2}}
\hspace{10pt} , \hspace{10pt}
0\le \; e \; < \; \frac{1}{2}
\hspace{10pt} .
\end{equation}
Larger values indicate a larger degree of elongation for all three measures.
There are, however, further possibilities to quantify structures,
see Section \ref{Elongation_HS_CS} and the Appendix.


\section{Cosmological Dependence of Gaussian Random CMB Maps}
\label{Cosmological_Dependence_of_Gaussian_random_CMB_map}



\subsection{Number of maxima and minima}
\label{Number_of_Maxima_and_Minima}


Before we turn to the elongation at maxima and minima,
we will discuss the number $N_{\hbox{\scriptsize max}}$ and 
$N_{\hbox{\scriptsize min}}$
of maxima and minima per solid angle on a 2-sphere.
Assuming that the temperature fluctuations
$\delta T\left(\hat{n} \right)$ 
are given by a homogeneous and isotropic Gaussian random field,
the corresponding derivations of the analytical formulae for the
ensemble average of the  number of maxima and minima per solid 
angle on a 2-sphere are given to some extent in
Ref.~\refcite{Bond_Efstathiou_1987}, 
and further relations can be found in \ref{appendix_peak_density}.

In the case of a homogeneous and isotropic universe the distribution of 
$N_{\hbox{\scriptsize max}}$ as a function of the temperature threshold
$\delta T$
is determined completely by the three parameters $\sigma^2_0$, $\sigma^2_1$
and $\sigma^2_2$, which are given by
\begin{equation}
\label{Eq:sigma_n}
\sigma_n^2 \; = \;
\sum_l\frac{2l+1}{4\pi} \, C_l \, |F_l|^2 \,
\frac{\left(l+n\right)!}{\left(l-n\right)!}
\hspace{10pt} \hbox{with} \hspace{10pt}
n \, = \, 0, 1, 2
\hspace{10pt} ,
\end{equation}
as discussed in the Appendix
(see (\ref{sigma_0}), (\ref{sigma_1}) and (\ref{sigma_2})).
Obviously these parameters can be calculated from the power spectrum
of the CMB. 
In this subsection we consider the number $N_{\hbox{\scriptsize max}}$
of maxima per solid angle depending on the normalised temperature
$\nu:=\frac{\delta T}{\sigma_0}$
where $\sigma_0$ is the standard deviation of the temperature fluctuations. 
The corresponding number of minima is given by the relation 
$N_{\hbox{\scriptsize min}}(\nu)=N_{\hbox{\scriptsize max}}(-\nu)$.

\begin{figure}[htb]
\begin{center}
\hspace*{-50pt}\begin{minipage}{18.5cm}
\vspace*{-20pt}
\includegraphics[width=9.5cm]{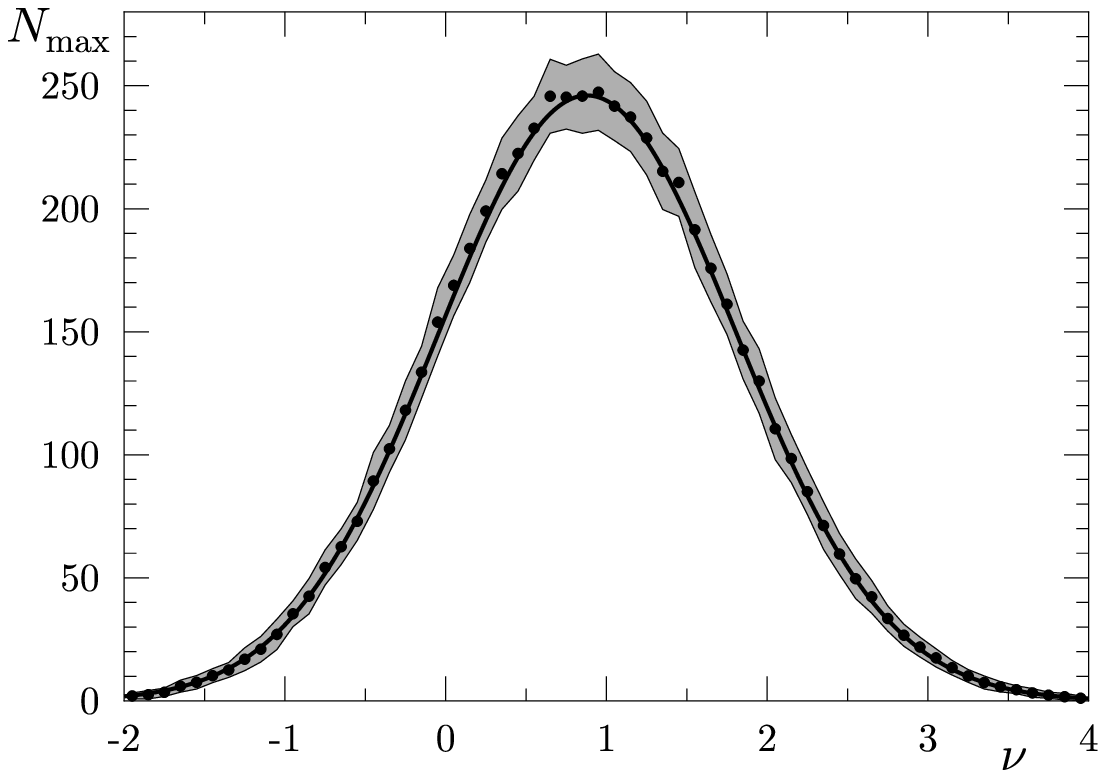}
\hspace*{-45pt}
\includegraphics[width=9.5cm]{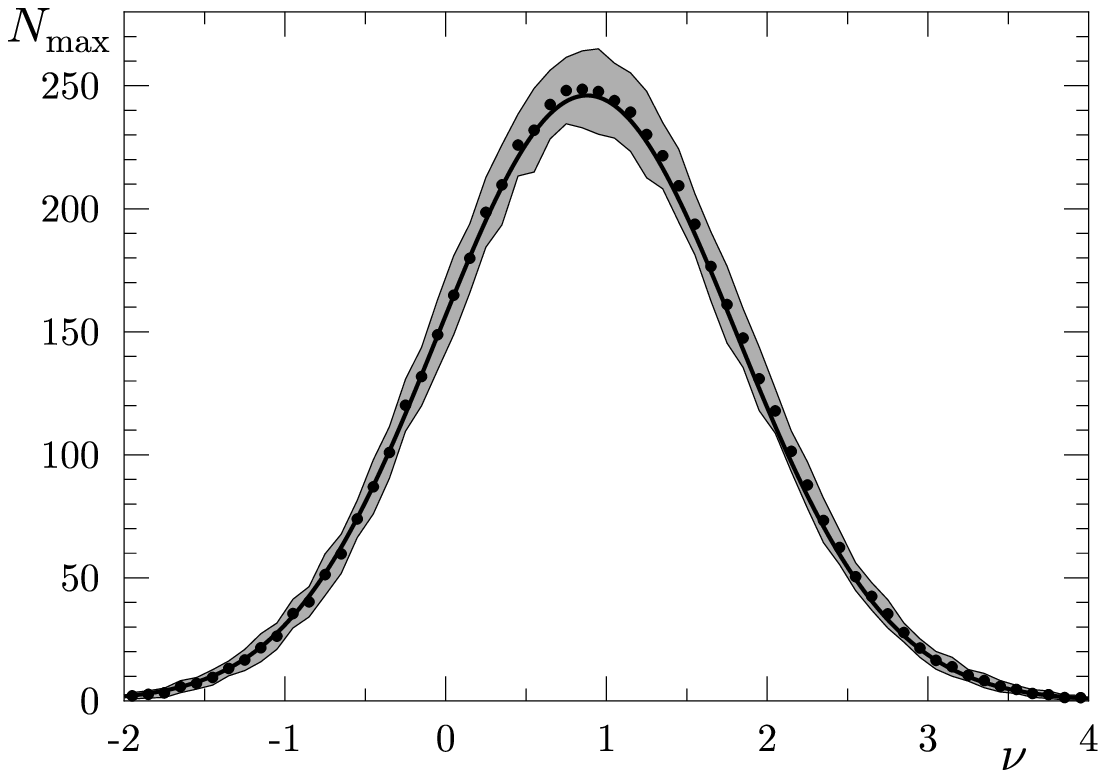}
\end{minipage}
\put(-470,45){(a)}
\put(-235,45){(b)}
\vspace*{-20pt}
\end{center}
\caption{\label{Fig:N_max_nu_torus}
The ensemble average of $N_{\hbox{\scriptsize max}}(\nu)$
for the $\Lambda$CDM concordance model
computed by Eq.~(\ref{Eq:N_max_nu_ht}) is shown as a solid line.
Panel (a) compares this curve with the average of
$N_{\hbox{\scriptsize max}}(\nu)$ computed from 50 realisations
of the infinite volume model (circles).
Panel (b) shows the result obtained from 50 realisations of the torus universe
with the side length $L=3.86\,L_H$ (circles).
To reveal the cosmic variance,
the $1\sigma$ standard deviation is displayed as a grey band
which is calculated from the corresponding 50 maps.
}
\end{figure}

For a homogeneous and isotropic Gaussian random field,
the distribution $N_{\hbox{\scriptsize max}}(\nu)$ is given by 
\begin{equation}
\label{Eq:N_max_nu_ht}
N_{\hbox{\scriptsize max}}(\nu) = \frac{1}{(2\pi)^\frac{3}{2}\,\theta^{*2}}
\exp\left[-\frac{\nu^2}{2}\right] \,G(\nu,\gamma,\alpha)
\hspace{5pt}
\end{equation}
with
\begin{eqnarray} \nonumber
G(\nu,\gamma,\alpha)&:=&\gamma \nu (1-\gamma^2) \frac{\exp\left[-\frac{\gamma^2\,\nu^2}
{2(1-\gamma^2)}\right]}{\sqrt{2\,\pi\,(1-\gamma^2)}}
\\ & & \hspace{0pt}\nonumber
+\left(\alpha^2(1-\gamma^2)-1+\gamma^2\nu^2\right)
\left[1-\frac{1}{2}\hbox{erfc}
\left( \frac{\gamma  
\nu}{\sqrt{2(1-\gamma^2)}}\right) \right]
\\ & & \hspace{0pt}
\nonumber
+\frac{\exp\left[\frac{-\alpha^2\,\gamma^2\,\nu^2}
{1+2\alpha^2(1-\gamma^2)}\right]}{\sqrt{2\alpha^2(1-\gamma^2)+1}}
\left[1-\frac{1}{2}\hbox{erfc}
\left( \frac{\gamma  
\nu}{\sqrt{2(1-\gamma^2)(1+2\alpha^2(1-\gamma^2))}}\right)\right]
\hspace{5pt},
\end{eqnarray}
and is derived in \ref{appendix_peak_density},
see Eq.~(\ref{N_max_nu}).
Here $\hbox{erfc}(x)$ is the complementary error function.
The parameters are
\begin{equation}
\label{Eq:alpha}
\alpha:=\sqrt{1+\frac{2\sigma^2_1}{\sigma^2_2}}
\hspace*{10pt} , \hspace*{10pt}
\theta^{*2}:=\alpha^2-1=\frac{2\sigma^2_1}{\sigma^2_2}
\hspace*{10pt} \hbox{, \; and} \hspace*{10pt}
\gamma:=\frac{\sigma_1^2}{\sigma_2\,\sigma_0\,\alpha}
\hspace*{10pt} .
\end{equation}
In Fig.~\ref{Fig:N_max_nu_torus} the theoretical distribution
(\ref{Eq:N_max_nu_ht}) is shown and compared with the mean value
obtained from 50 CMB maps
simulated at the HEALPix resolution $N_{\hbox{\scriptsize side}}=512$
using a smoothing of $1^{\circ}$ and $l \le 1000$.
Both panels are based on the cosmological parameters of the infinite
volume best-fit $\Lambda$CDM model of the WMAP data.
In panel (a) the infinite model, which is isotropic, is shown,
whereas panel (b) shows a multi-connected model which is anisotropic
and thus does not fulfil the assumptions on which (\ref{Eq:N_max_nu_ht})
is based.
However, as revealed by Fig.~\ref{Fig:N_max_nu_torus}(b)
the data points match the distribution (\ref{Eq:N_max_nu_ht}) with the same
quality as in Fig.~\ref{Fig:N_max_nu_torus}(a).
In addition, the 50 realisations can be used to get an estimate of the
cosmic variance.
The corresponding $1\sigma$ standard deviation is displayed as a grey band
in Fig.~\ref{Fig:N_max_nu_torus}.
The results for the mean value and the variance of
$N_{\hbox{\scriptsize max}}(\nu)$ for the torus universe are
almost the same as for the infinite one. 
One could have expected this result
since the number of maxima is primarily determined
by the smallest scales, and a multi-connected model,
which possesses in the shown case a cubic topology with
a side length $L=3.86\,L_H$,
differs from the infinite one only on very large scales.
Similar results are also expected in the case of other topologies, 
homogeneous and inhomogeneous manifolds,
if the topological scales of the universe are of comparable order.
Only if the volume of the universe is much smaller than the volume
inside the surface of last scattering, 
one can expect any differences in the $N_{\hbox{\scriptsize max}}$
distribution to those calculated in \ref{appendix_peak_density}. 
In such a case one would get an influence of the topology on small scales.
But until now no hint is found in the data for a model with such
a sufficiently small volume.

\begin{figure}[htb]
\begin{center}
\hspace*{-50pt}\begin{minipage}{18.5cm}
\vspace*{-20pt}
\includegraphics[width=9.5cm]{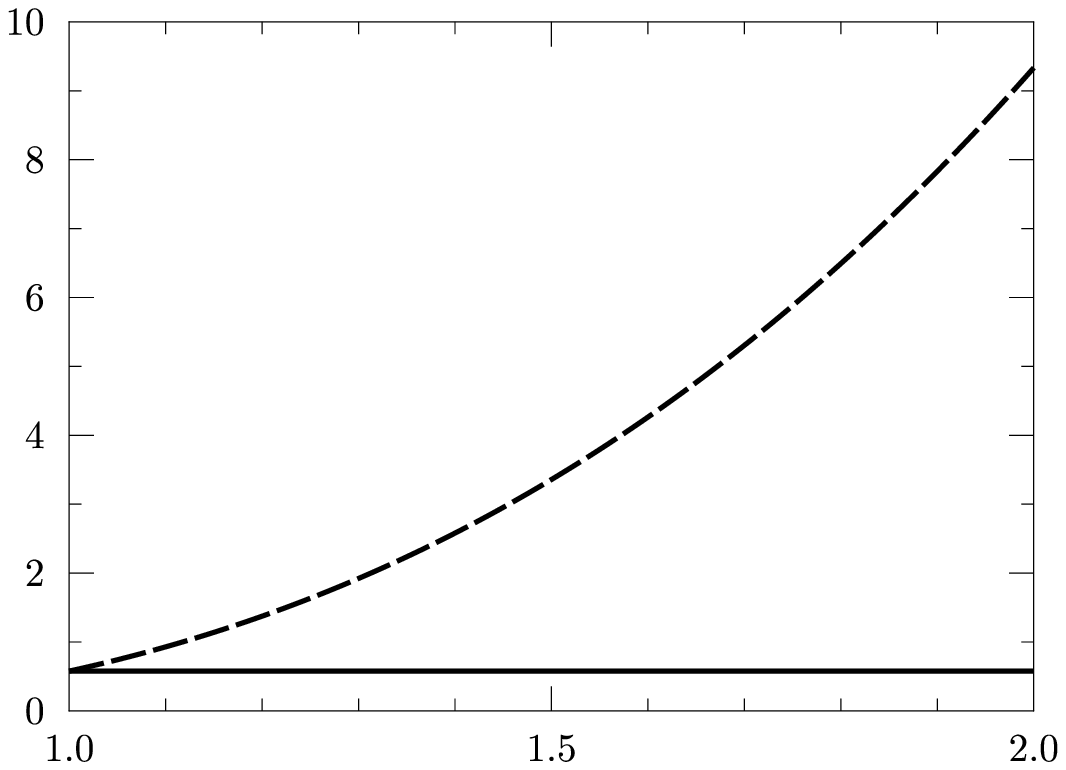}
\hspace*{-45pt}
\includegraphics[width=9.5cm]{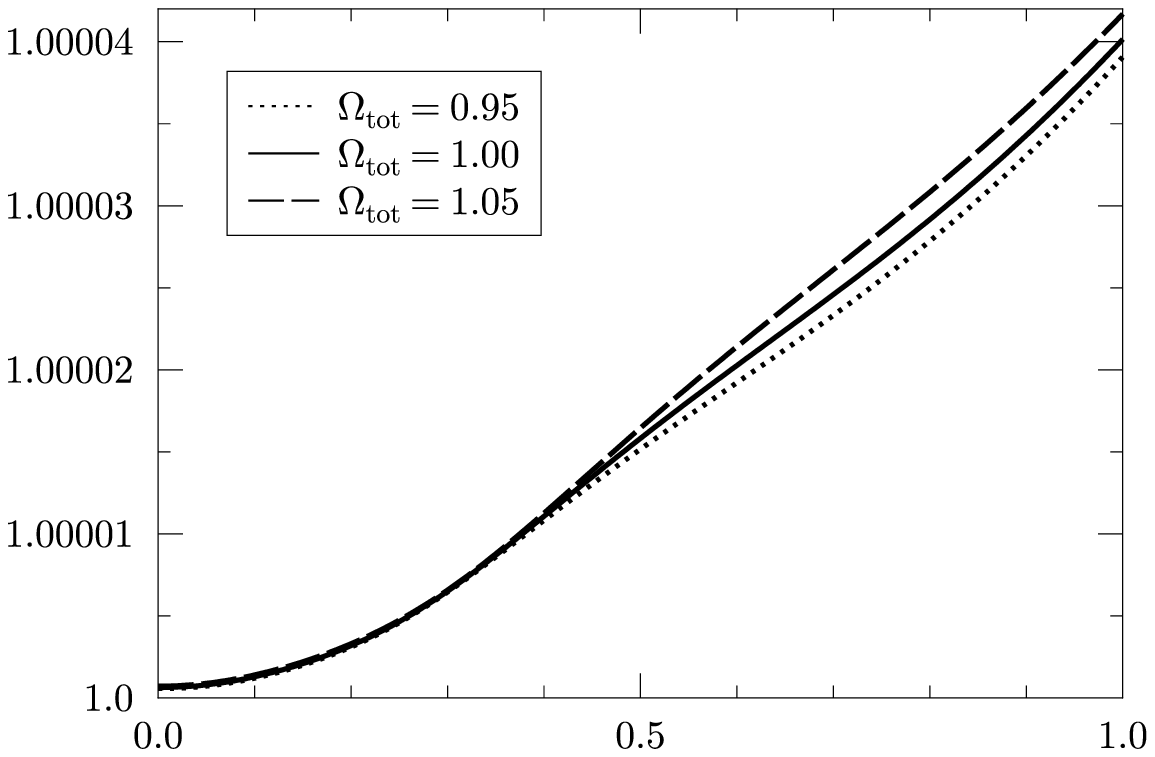}
\end{minipage}
\put(-470,-35){(a)}
\put(-510,5){\begin{turn}{90}
\small $4\pi\theta^{*2}\,N_{\hbox{\scriptsize max}}$
\end{turn}}
\put(-320,-70){$\alpha$}
\put(-235,-35){(b)}
\put(-270,55){$\alpha$}
\put(-124,-70){$\sigma_{\hbox{\scriptsize fwhm}}$ [deg]}
\vspace*{-20pt}
\end{center}
\caption{\label{Fig:N_max_analytic_dep_a}
In panel (a) the formula of the total number $N_{\hbox{\scriptsize max}}$
of maxima per solid angle resulting from Equation (A1.7)
in Ref.~\protect\refcite{Bond_Efstathiou_1987}
(solid line) is compared to our Eq.~(\ref{Eq:N_max_ht}) (dashed line).
In panel (b) the parameter $\alpha$ is plotted as a function
of the smoothing scale $\sigma_{\hbox{\scriptsize fwhm}}$
for three cosmological models.
All values of $\alpha$ are very close to one.
}
\end{figure}

Our formulae for the various $N_{\hbox{\scriptsize max}}$ distributions
in \ref{appendix_peak_density} differ by terms depending on the parameter
$\alpha$ from the corresponding equations
in Ref.~\refcite{Bond_Efstathiou_1987}. 
Here these differences are discussed using the example of the
total number of maxima per solid angle. 
Our formula to this quantity is given by 
\begin{eqnarray} 
\label{Eq:N_max_ht}
N_{\hbox{\scriptsize max}}& = &
N_{\hbox{\scriptsize max}}^{\hbox{\scriptsize B\&E}}\,
\frac{\left[\left(\alpha^{2}-1\right)\sqrt{1+2\alpha^{2}}+1\right]\sqrt{3}}
{\sqrt{1+2\alpha^{2}}}
\hspace{5pt}
\end{eqnarray}
which is derived in \ref{appendix_peak_density}, see (\ref{N_max}).
The corresponding result obtained from Equation (A1.7)
in Ref.~\refcite{Bond_Efstathiou_1987} is
$N_{\hbox{\scriptsize max}}^{\hbox{\scriptsize B\&E}}=
\frac{1}{4\pi\theta^{*2}\sqrt{3}}$,
which is the leading term of the  Laurent series of 
our Eq.~(\ref{Eq:N_max_ht}) at $\alpha^2=1$.
Both formulae are compared in Fig.~\ref{Fig:N_max_analytic_dep_a}(a)
where $4\pi\theta^{*2}\,N_{\hbox{\scriptsize max}}$ is displayed.
In the case of $N_{\hbox{\scriptsize max}}^{\hbox{\scriptsize B\&E}}$
an $\alpha$-independent,
i.\,e.\ model independent value of $\frac{1}{\sqrt{3}}$ is obtained.
In contrast our result for $4\pi\theta^{*2}\,N_{\hbox{\scriptsize max}}$
depends on $\alpha$ and therefore on the cosmological model
since $\alpha$, $\sigma_1^2$ and $\sigma_2^2$
depend on the angular power spectrum.
In Fig.~\ref{Fig:N_max_analytic_dep_a}(a) we have varied $\alpha\in [1,2]$
in order to demonstrate the difference between
$N_{\hbox{\scriptsize max}}$ and the leading term
$N_{\hbox{\scriptsize max}}^{\hbox{\scriptsize B\&E}}$.
Is this interval of $\alpha$ realistic for cosmological models
compatible with the measured data?
To answer this question we have calculated $\sigma_1^2$ and $\sigma_2^2$
with a cut-off at $l=1000$ with the angular power spectrum of the
best-fit $\Lambda$CDM model of the WMAP data. 
In addition, these parameters are computed from the angular power spectrum
of two models with positive and negative curvature,
where the cosmological constant is varied and
the other cosmological parameters are held fixed.
These three cosmological models lead to different values of $\alpha$,
whose dependence on the smoothing scale is displayed in
Fig.~\ref{Fig:N_max_analytic_dep_a}(b).
One observes that $\alpha$ is almost equal to 1, and thus
$\theta^{*2}$ is very small 
for all realistic cosmological parameters and smoothing scales.
Therefore, we conclude that for all practical purposes
the additional dependence of
$4\pi\theta^{*2}\,N_{\hbox{\scriptsize max}}$
on $\alpha$ is without relevance and the formulae of the
$N_{\hbox{\scriptsize max}}$ distributions
in Ref.~\refcite{Bond_Efstathiou_1987} are adequate.


\subsection{Elongation at Maxima and Minima} 
\label{Ellipticity_at_Maxima_and_Minima}

\begin{figure}[htb]
\begin{center}
\hspace*{-50pt}\begin{minipage}{18.5cm}
\vspace*{-20pt}
\includegraphics[width=9.5cm]{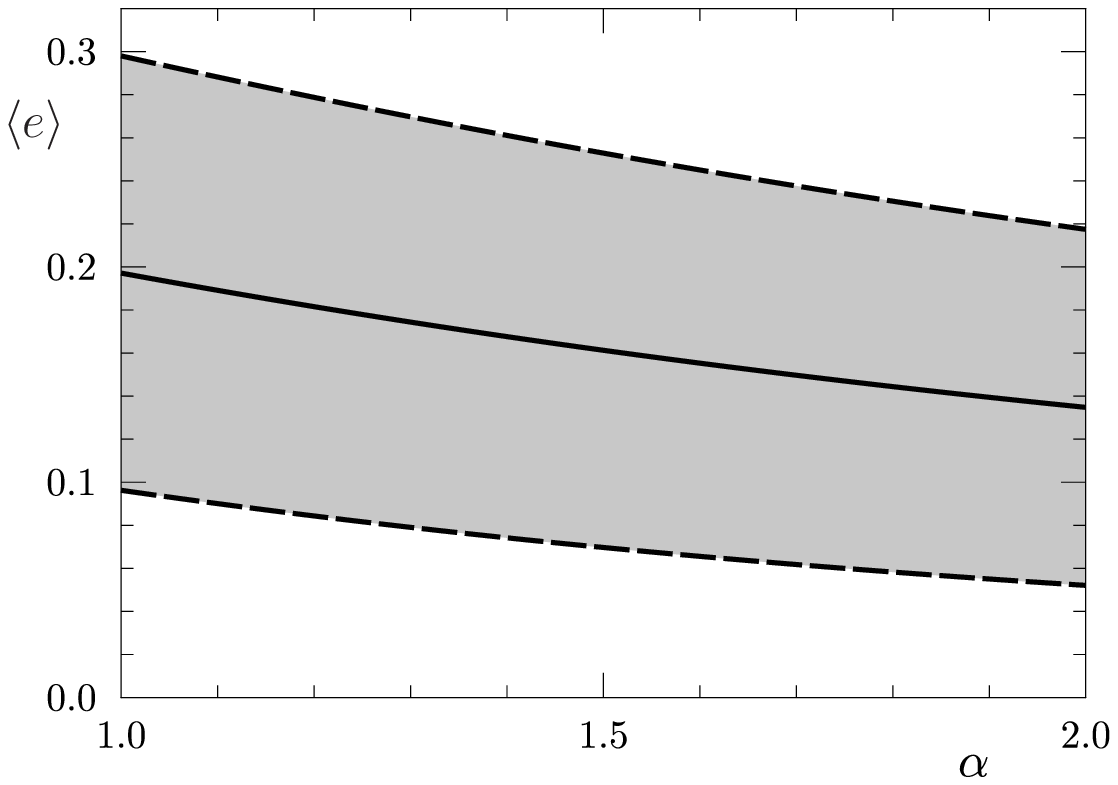}
\hspace*{-45pt}
\includegraphics[width=9.5cm]{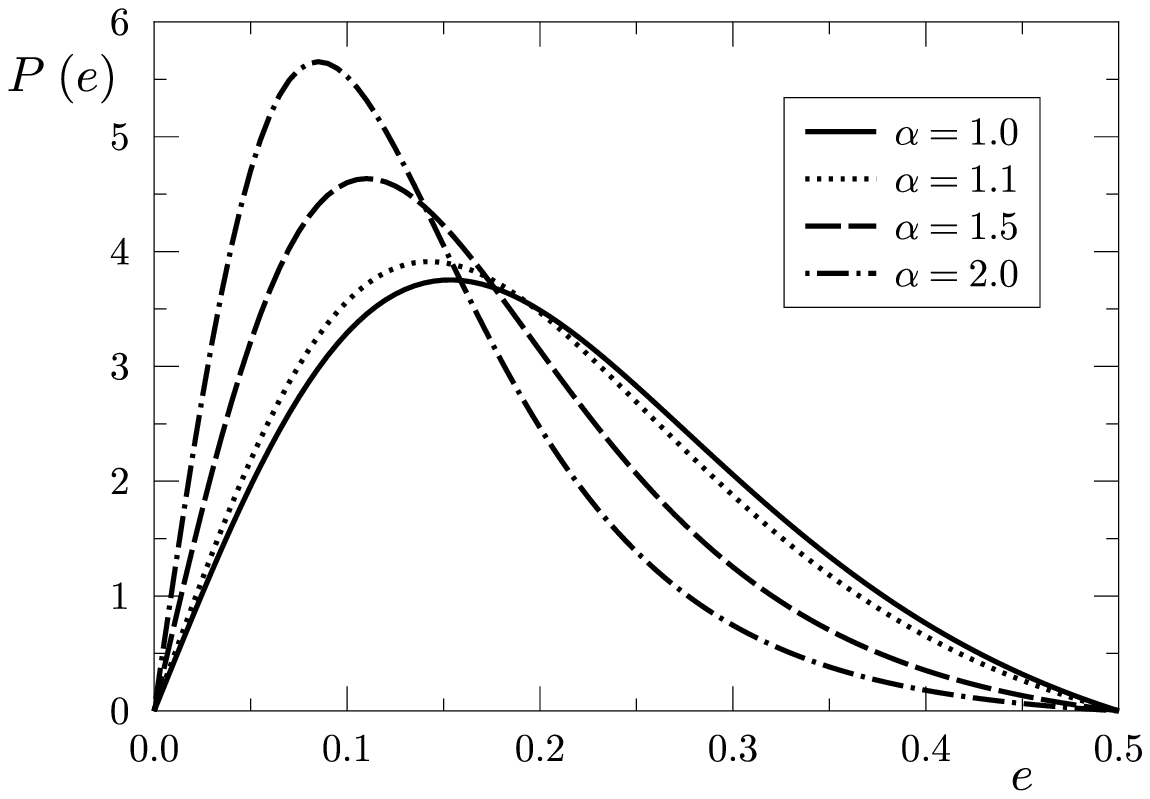}
\end{minipage}
\put(-470,-35){(a)}
\put(-235,-35){(b)}
\vspace*{-20pt}
\end{center}
\caption{\label{Fig:Distribution_ellipticiy_analytic_dep_a}
In panel (a) the ensemble average of the elongation $e$ with the corresponding
$1\sigma$ standard deviation is displayed depending on the parameter $\alpha$.
In panel (b) the distribution $P(e)$ of the elongation $e$ is plotted
depending on the parameter $\alpha$.
}
\end{figure}

Now we turn to the elongation measure $e$ at maxima and minima.
The ensemble average and the corresponding second moment of this quantity
are given by 
\begin{eqnarray}
\label{Eq:mittelwert_e_ht}
\hspace*{-15pt}\langle e \rangle
\; = \;
\frac{2\alpha^2+3-\frac{3}{2}\sqrt{\frac{1+2\alpha^2}{2\alpha^2}}  \ln
\left[\frac{1+\sqrt{\frac{2\alpha^2}{1+2\alpha^2}}}{1-\sqrt{\frac{2\alpha^2}
{1+2\alpha^2}}}\right]}{4 \left[ (\alpha^2-1) \sqrt{1+2\alpha^2} +1
\right]}
\hspace{5pt}
\end{eqnarray}
and
\begin{eqnarray}
\label{Eq:zweite_moment_e_ht}
\langle e^2 \rangle
\;=\; \frac{1-3\,\sqrt{1+2\alpha^2} + 4
(\sqrt{1+2\alpha^2}-1)\left(\frac{1+2\alpha^2}{2\alpha^2}\right)}{4 \left[ (\alpha^2-1)
\sqrt{1+2\alpha^2} +1 \right]}
\hspace{5pt} ,
\end{eqnarray}
which both depend on the parameter $\alpha$. 
The formulae are derived in the \ref{appendix_Ellipticity_of_the_CMB}.
The mean value of the elongation $e$ together with the standard deviation
due to the cosmic variance is displayed in panel (a) of
Fig.~\ref{Fig:Distribution_ellipticiy_analytic_dep_a}.
The distribution
\begin{eqnarray}
\label{Eq:P_e_ht}
P(e) \;=\; \frac{24\, e (1-4e^2)\alpha^4 \sqrt{1+2\alpha^2}}{\left(1+8\alpha^2e^2\right)^{\frac{5}{2}}
\left[(\alpha^2-1)\sqrt{1+2\alpha^2}+1\right]} 
\hspace{5pt}
\end{eqnarray}
of the elongation is plotted for four values of $\alpha$ in
Fig.~\ref{Fig:Distribution_ellipticiy_analytic_dep_a}(b).
This demonstrates that $P(e)$ depends on the parameter $\alpha$
and thus in turn via the angular power spectrum
on the cosmological parameters.
In this way the elongation measure $e$ encodes some properties
of the Universe.

This dependence of the elongation $e$ on cosmological parameters 
is not discussed in Ref.~\refcite{Bond_Efstathiou_1987}. 
In currently admissible cosmological models the dependence on $\alpha$
in (\ref{mittelwert_e}) can be neglected
since the corresponding values of $\alpha$ are very close to one.
The ensemble average of the elongation $e$ is approximatively
given by 0.197 for $\alpha=1$.
Similar results are obtained for the other elongation measures,
i.\,e.\ for the ellipticity $E$ and the eccentricity $\varepsilon$.
The ensemble averages of these quantities are given by
(\ref{erwartungswert_E}) and (\ref{erwartungswert_epsilon}).
Their values are approximately $E \approx 1.648$ and
$\varepsilon \approx 0.715$ independent from the model.  
The cosmological influence on the elongation $e$,
the ellipticity $E$ or the eccentricity $\varepsilon$
is very difficult to measure.
The difficulty arises from the large cosmic variance for these quantities
which is displayed as a grey band in 
Fig.~\ref{Fig:Distribution_ellipticiy_analytic_dep_a}(a)
in the case of the elongation $e$. 
Furthermore, also the detector noise and the foregrounds in a sky map
obtained by observations have an influence on the elongation,
but this will be discussed later.

In addition, it is not expected that the topology of the Universe,
i.\,e.\ a multi-connected spatial space,
can influence the elongation of contour lines
since the volumes of the fundamental cells are too large.
Nevertheless, several measures of the elongation are discussed below
for the special case of the cubic topology
which is motivated by the fact that this model is statistically anisotropic.


\section{Elongation of Hot and Cold Spots}
\label{Elongation_HS_CS}


There are different algorithms to compute the ellipticity.
One method is based on the Taylor expansion of the temperature field
at local maxima or minima as used in the Appendix.
Alternatively, the ellipticity can be calculated by using an
``inertia'' tensor as described in the following.

\subsection{Computation of the ellipticity}

In the following the normalised temperature field
\begin{equation}
\label{equ:norm}
u(\hat{n}) \; = \; \frac{\delta T_{F_l} (\hat{n})}{\sigma_0}
\end{equation}
is used.
In the theoretical derivation which is outlined in the Appendix,
this $\sigma_0$ is obtained by the ensemble average
by using $\sigma_0 = \sqrt{C(0)}$
where $C(0)$ is the 2-point correlation function
of the CMB at a separation angle $\vartheta=0$.
Such an ensemble average is not possible for a CMB sky map
obtained by observations, of course,
since there is only a single CMB sky from our point of view.
Therefore, a sky map analysis computes the value of $\sigma_0$
from a given single sky realisation as the usual standard deviation
of the temperature field
\begin{equation}
\label{equ:norm_sigma}
\sigma_0 = \sqrt{N_{\delta \Omega}^{-1} \int_{\delta \Omega}
(\delta T_{F_l} (\hat{n}) - \bar{T})^2 d \Omega}
\hspace*{10pt} ,
\end{equation}
where $\bar{T}$ is the mean temperature.
In the analysis of masked maps the observed area is $\delta \Omega$
and $N_{\delta \Omega}$ is the fraction of the sky that is not masked.

To get a binary image we define the excursion set of hot spots (HS) by
\begin{equation}
\label{equ:exc_h}
Q_{\nu}^{\hbox{\scriptsize HS}} \; =\;
\left\{ \hat{n} \in \mathbb{S}^2 | u(\hat{n}) \geq \nu \right\}
\hspace*{10pt} .
\end{equation}
Here one excludes all regions with a temperature value lower than the value
of the threshold $\nu$.
Analogously the excursion set of cold spots (CS) is defined by
\begin{equation}
\label{equ:exc_cs}
Q_{\nu}^{\hbox{\scriptsize CS}} \; =\;
\left\{ \hat{n} \in \mathbb{S}^2 | u(\hat{n}) \leq \nu \right\}
\hspace*{10pt} .
\end{equation}
The multi-connected set $Q_{\nu}$ can be interpreted as the union
\begin{equation}
\label{equ:exc_union}
Q_{\nu} \; = \;\bigcup\limits_{i=1}^{N_{\nu}} Q_{\nu}^{i}
\end{equation}
of the $N_{\nu}$ simply connected spots $Q_{\nu}^{i}$.
We compute the ellipticity for every spot $Q_{\nu}^{i}$
by using the inertia tensor defined by
\begin{equation}
I(Q_{\nu}^{i}) \; := \;
\left( \begin{tabular}{rr}
$I_{xx}$ & $I_{xy}$ \\
$I_{yx}$ & $I_{yy}$
\end{tabular} \right)
\; = \;
\left( \begin{tabular}{rr}
$\int\int y^2 dxdy$ & $-\int\int yx\, dxdy$ \\
$-\int\int xy\, dxdy$ & $\int\int x^2 dxdy$
\end{tabular} \right)
\hspace*{10pt} .
\end{equation}
The spots are projected onto the $xy$-plane
where the origin of the coordinate system matches the centre of mass of
the spot $Q_{\nu}^{i}$.
The symmetric tensor $I(Q_{\nu}^{i})$ has two real eigenvalues
$\lambda_1$ and $\lambda_2$
whose relation to the ellipticity $E$ is given by
\begin{equation}
\label{EQ:EQ}
E(Q_{\nu}^{i}) \; = \; \sqrt{\frac{\lambda_1}{\lambda_2}}
\hspace{10pt} .
\end{equation}
For the special case of an ellipse with axes $a$ and $b$
the eigenvalues can be computed to be
\begin{equation}
\lambda_1 \; = \; \frac\pi 4 \,a^3 \, b
\hspace{10pt} \hbox{ and } \hspace{10pt}
\lambda_2 \; = \; \frac\pi 4 \, a \, b^3
\hspace{10pt} \hbox{ with } \hspace{10pt}
\lambda_1 \geq \lambda_2
\hspace{10pt} .
\end{equation}
Inserting this into (\ref{EQ:EQ}) leads back to the definition
(\ref{Eq:Ellipticity}), respectively (\ref{Eq:def_ellip})
of the ellipticity
\begin{equation}
E \; = \; \frac{a}{b} \; = \; \sqrt{\frac{\lambda_1}{\lambda_2}}
\hspace{10pt} .
\end{equation}
The other quantities for elongation given in (\ref{Eq:def_excen}) and
(\ref{Eq:def_e}) can be obtained from the conversion formulae.
The true niveau lines are not perfect ellipses, of course,
but the above prescription can be used in order to obtain
the ellipticity of that ellipse which approximates best the niveau line.
Since there are niveau lines which deviate strongly from an ellipse
one needs a further criterion to eliminate such curves in order
to avoid non-sense results.


\subsection{The dependence of the ellipticity on the resolution}


The ellipticity depends sensitively on the accuracy of the sky map
and care has to be taken with respect to the resolution of the map,
the beam profile and the noise properties.
In so far, it is not a very robust measure to characterise the CMB sky,
and the details of the computation of the ellipticity
have to be defined clearly in order to obtain reproducible values.

\begin{figure}[htb]
\begin{center}
\hspace*{-10pt}\begin{minipage}{12cm}
\vspace*{-25pt}
\includegraphics[width=12.0cm]{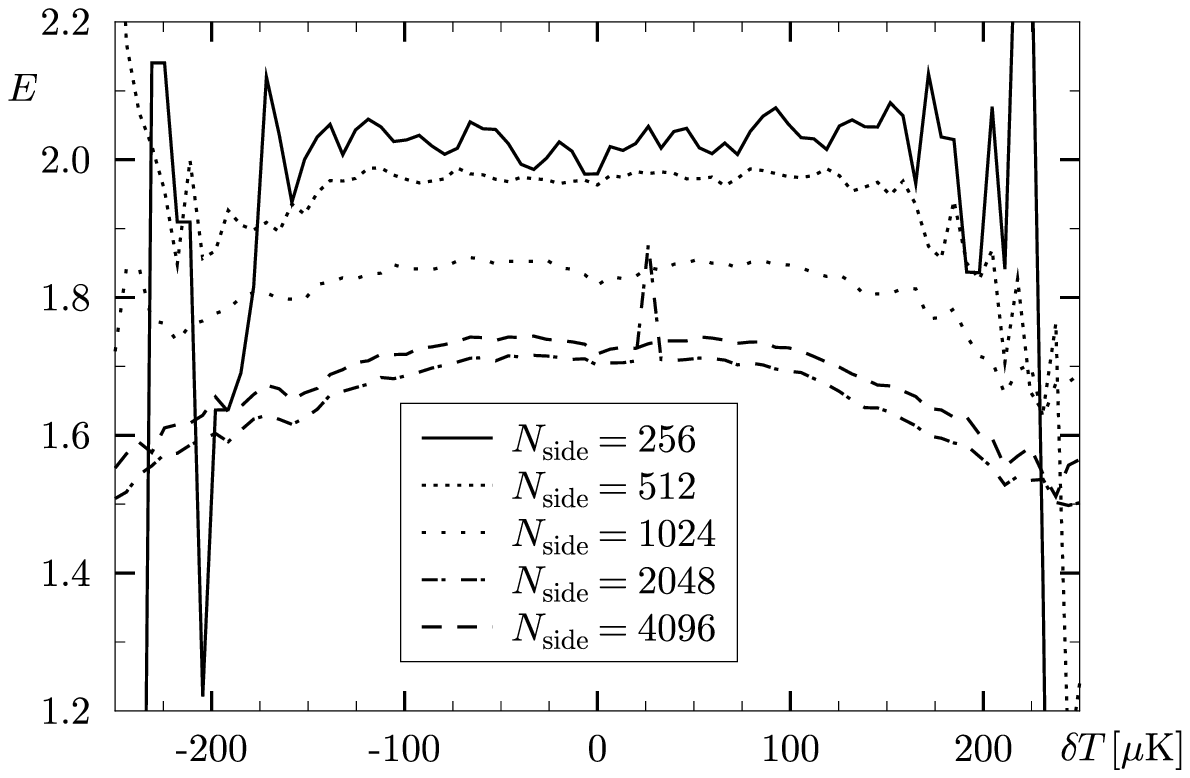}\vspace*{-25pt}
\includegraphics[width=12.0cm]{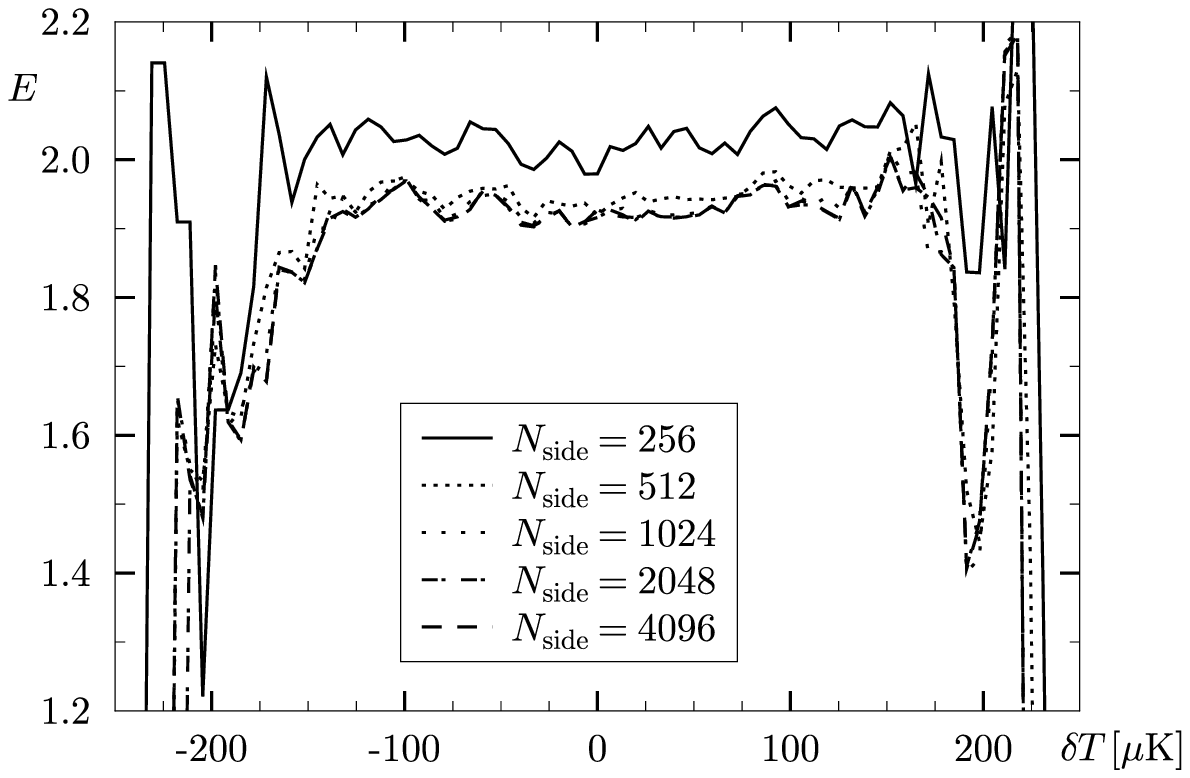}
\end{minipage}
\put(-235,195){(a)}
\put(-235,-25){(b)}
\vspace*{-20pt}
\end{center}
\caption{\label{Fig:Ellipticiy_Healpix_nside}
The ellipticity $E$ is computed for a realisation
of the cubic universe as described in the text
based on the cosmological parameters of the concordance model.
The dependence on the value of $N_{\hbox{\scriptsize side}}$ is presented.
Panel (a) shows the ellipticity $E$ where all contours are taken
into account that enclose at least 10 pixels.
The ellipticity $E$ increases with decreasing $N_{\hbox{\scriptsize side}}$.
In Panel (b) all contours are selected enclosing
an area of at least $0.525\,\hbox{deg}^2$.
}
\end{figure}

Let us at first address the issue of the resolution of the sky map.
The CMB sky maps are usually stored in the
HEALPix\cite{Gorski_Hivon_Banday_Wandelt_Hansen_Reinecke_Bartelmann_2005}
format
whose resolution is determined by the parameter $N_{\hbox{\scriptsize side}}$.
In order to determine the dependence of the ellipticity $E$ on the
value of $N_{\hbox{\scriptsize side}}$,
a sky map is simulated for the cubic topology using the
cosmological parameters of the $\Lambda$CDM concordance model.
We start with a highest resolution of $N_{\hbox{\scriptsize side}}=4096$.
The sky maps are normalised such that the temperature fluctuations
possess a standard deviation of $66 \mu\hbox{K}$,
which is the value obtained from the five-year ILC map outside the KQ75 mask.

For such a simulation the ellipticity $E$ is computed for
the contour lines of a given temperature threshold $\delta T$,
and the mean value is shown in Fig.~\ref{Fig:Ellipticiy_Healpix_nside}.
Contour lines that are too small are excluded
since they are determined by very few pixels
such that the ellipticity $E$ is ill defined.
In Fig.~\ref{Fig:Ellipticiy_Healpix_nside}(a) the criterion is
that the contour lines should at least circumference 10 pixels
with respect to the chosen resolution $N_{\hbox{\scriptsize side}}$.
This implies that ever more small contour lines are taken into account
as the value of $N_{\hbox{\scriptsize side}}$ increases.
Thus the mean values are computed from different sets of contour lines.
This contrasts to the criterion of a fixed minimal area
which leads to the result displayed in
Fig.~\ref{Fig:Ellipticiy_Healpix_nside}(b).
There, all contour lines are taken into account
which are larger than 10 pixels with respect to the
$N_{\hbox{\scriptsize side}}=256$ resolution,
i.\,e.\ all contours encompassing an area larger than $0.525\,\hbox{deg}^2$.
Thus the curve belonging to $N_{\hbox{\scriptsize side}}=256$
is the same in both panels.
Whereas the exclusion criterion of a fixed pixel number leads
to a robust estimation of $E$ only at very large values of
$N_{\hbox{\scriptsize side}}$,
the exclusion criterion of a fixed area gives consistent values
already for values of $N_{\hbox{\scriptsize side}}$ as small as 512.
But recall from Section \ref{Ellipticity_at_Maxima_and_Minima}
that the expected ellipticity is $E\simeq 1.648$ for a Gaussian field.
This value is, however, obtained with the first selection criterion
which takes an increasing number of contour lines into account.
This analysis thus favours the first selection criterion and,
furthermore, demonstrates the sensitivity of the elongation measures.

\begin{figure}[htb]
\begin{center}
\hspace*{-10pt}\begin{minipage}{12cm}
\vspace*{-25pt}\includegraphics[width=12.0cm]{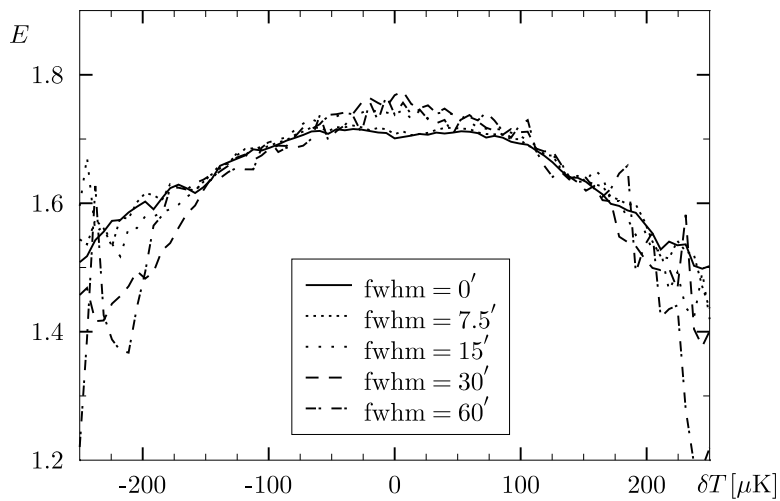}
\end{minipage}
\vspace*{-30pt}
\end{center}
\caption{\label{Fig:Ellipticiy_Healpix_fwhm}
For the CMB simulation used in Fig.~\ref{Fig:Ellipticiy_Healpix_nside}(a)
having the HEALPix resolution of $N_{\hbox{\scriptsize side}}=4096$,
the ellipticity $E$ is shown in dependence on the width
of the Gaussian beam.
}
\end{figure}

The dependence of the ellipticity on the smoothing is not very strong.
This can be inferred from Fig.~\ref{Fig:N_max_analytic_dep_a}(b)
where the value of $\alpha$ is plotted as a function of the
smoothing $\sigma_{\hbox{\scriptsize fwhm}}$.
Up to a smoothing of 1 degree all values of $\alpha$ are below 1.00004
and the expectation value for the ellipticity $E$,
Eq.~(\ref{erwartungswert_E}), is nearly constant for
these smoothings.
This behaviour is confirmed in Fig.~\ref{Fig:Ellipticiy_Healpix_fwhm}
where different smoothings are applied to the sky maps with a fixed
HEALPix resolution of $N_{\hbox{\scriptsize side}}=4096$.


\subsection{The dependence of the elongation on the curvature}


Let us now turn to the dependence of the elongation on the curvature which is
claimed\cite{Gurzadyan_Torres_1997,Gurzadyan_et_al_2003a,Gurzadyan_et_al_2003b,%
Gurzadyan_et_al_2004,Gurzadyan_et_al_2005,Gurzadya_et_al_2007}
to be sufficiently strong in order to reveal the curvature
of the Universe.
However, in Ref.~\refcite{Barreiro_Sanz_Martinez-Gonzalez_Cayon_Silk_1997}
no such ellipticity-curvature correlation is found.
Since the elongations depend on the parameter $\alpha$
and thus on the cosmological parameters, 
such a correlation could exist.
The discussion in the previous section has shown
that the elongation depends on the size of the niveau lines
which have to enclose at least $n_{\hbox {\scriptsize min}}$ pixels.
For the computation of the elongation
our algorithm requires at least an area greater than $4$ pixels.
The number of niveau lines which can be used in a statistic
depends on the normalised temperature $\nu$.
It is maximal at $\nu=\pm 1$
where the number $N_{\nu}$ of spots has its maximum.
The values of $n_{\hbox {\scriptsize min}}$ and
of the normalised temperature $\nu$ are required to specify the
mean ellipticity $\bar{E}(\nu,n_{\hbox{\scriptsize min}})$.

The ensemble average of the elongation and the corresponding cosmic variance
are computed from an ensemble of 1000 sky realisations.
We analyse the excursion sets $Q_{\nu}$ for hot spots with
thresholds $\nu>0$ and for cold spots with thresholds $\nu<0$.
We get for every threshold $\nu$ an amount of $N_{\nu}$ spots
with individual areas $n(Q_{\nu}^{i})$ and elongations $e(Q_{\nu}^{i})$.
The mean ellipticity
$\bar{E}(\nu, n_{\hbox {\scriptsize min}})$
is computed at $\nu=1$ by averaging the values of all spots with
$n>n_{\hbox {\scriptsize min}}$.
This method is used in Refs.~\refcite
{Gurzadyan_et_al_2003b,%
Gurzadyan_et_al_2004,Gurzadyan_et_al_2005,Gurzadya_et_al_2007}.
An alternative averaging is used for the mean elongation
$\bar{e}(\nu, n, \Delta n)$
which is computed as a moving average around a spot size $n$,
i.\,e.\ from spots having pixel numbers within the interval
$\left[ n - \Delta n, n + \Delta n\right]$.
Our examination shows that these mean values are very stable
with respect to the choice of the threshold $\nu$ in a range of
$\nu \in \left[ -2,2\right]$.

\begin{figure}[htp]
\begin{center}
\hspace*{-50pt}
{
\begin{minipage}{18.5cm}
\includegraphics[angle=-90,width=7.5cm]{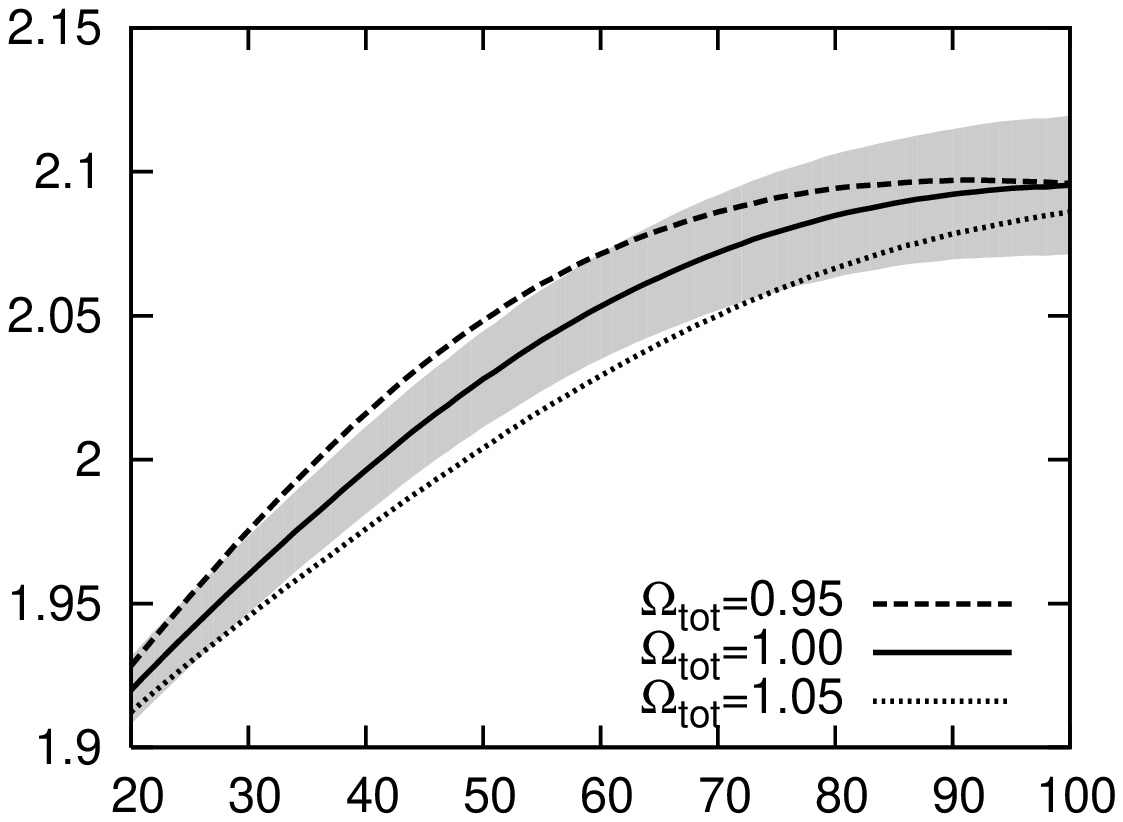}
\includegraphics[angle=-90,width=7.5cm]{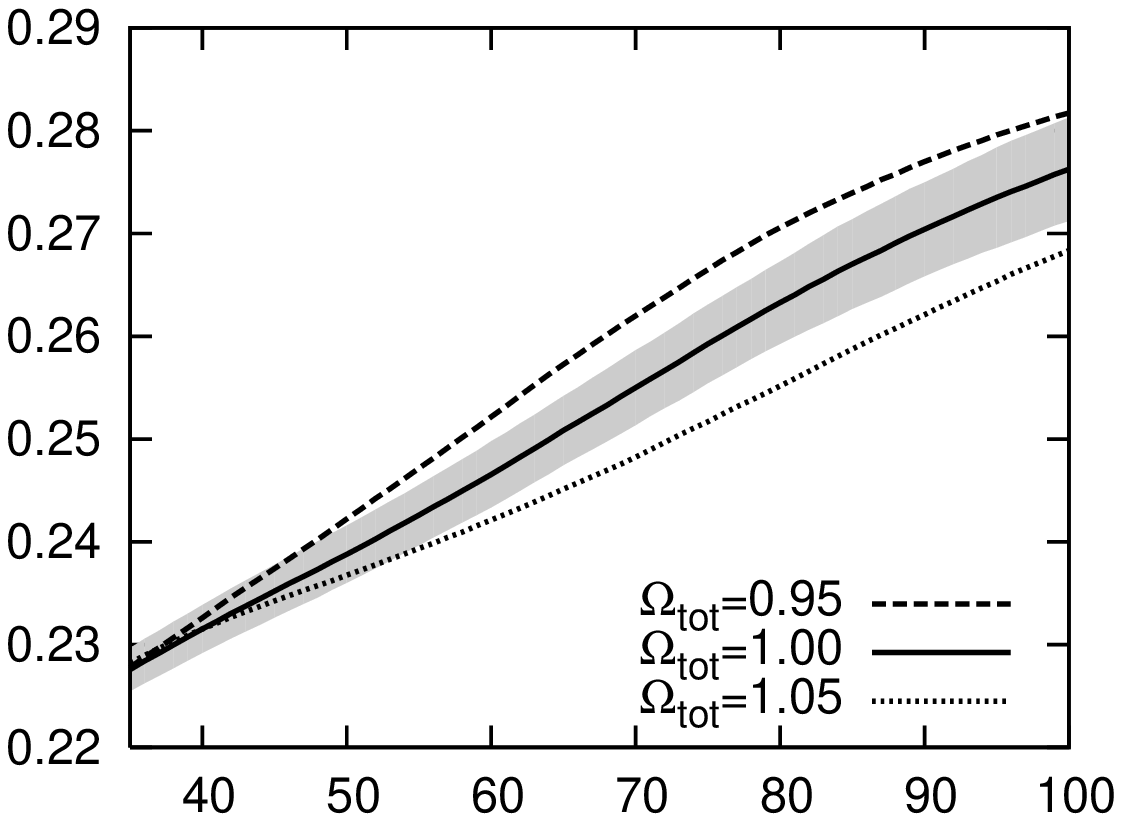}
\end{minipage}
\put(-480,45){ ${\sigma}_{\hbox{\scriptsize fwhm}} = 20\hbox{ arcmin}$}
\put(-260,45){ ${\sigma}_{\hbox{\scriptsize fwhm}} = 20\hbox{ arcmin}$}
\put(-515,52){$\bar{E}$}
\put(-340,-75){$n_{\hbox {\scriptsize min}}$}
\put(-300,55){$\bar{e}$}
\put(-125,-75){$n$}
}
\hspace*{-50pt}
{
\begin{minipage}{18.5cm}
\includegraphics[angle=-90,width=7.5cm]{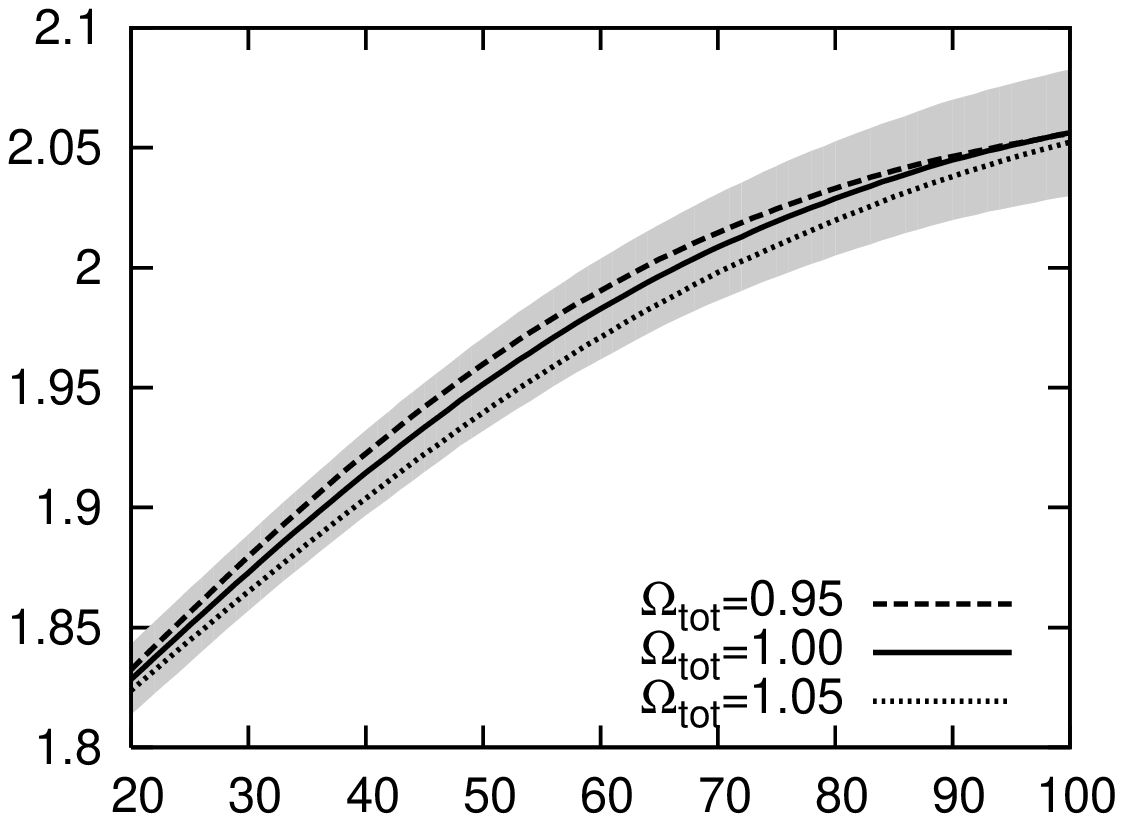}
\includegraphics[angle=-90,width=7.5cm]{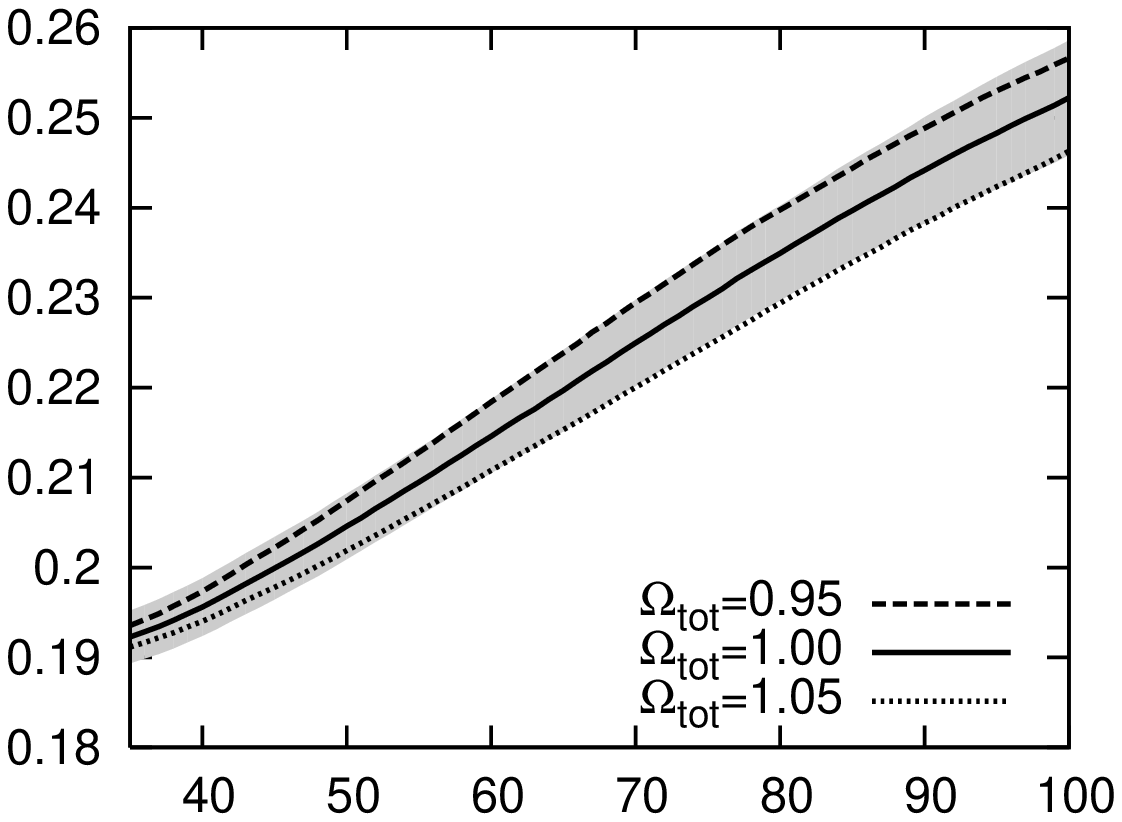}
\end{minipage}
\put(-480,45){ ${\sigma}_{\hbox{\scriptsize fwhm}} = 40\hbox{ arcmin}$}
\put(-260,45){ ${\sigma}_{\hbox{\scriptsize fwhm}} = 40\hbox{ arcmin}$}
\put(-515,52){$\bar{E}$}
\put(-340,-75){$n_{\hbox {\scriptsize min}}$}
\put(-300,55){$\bar{e}$}
\put(-125,-75){$n$}
}
\hspace*{-50pt}
{
\begin{minipage}{18.5cm}
\includegraphics[angle=-90,width=7.5cm]{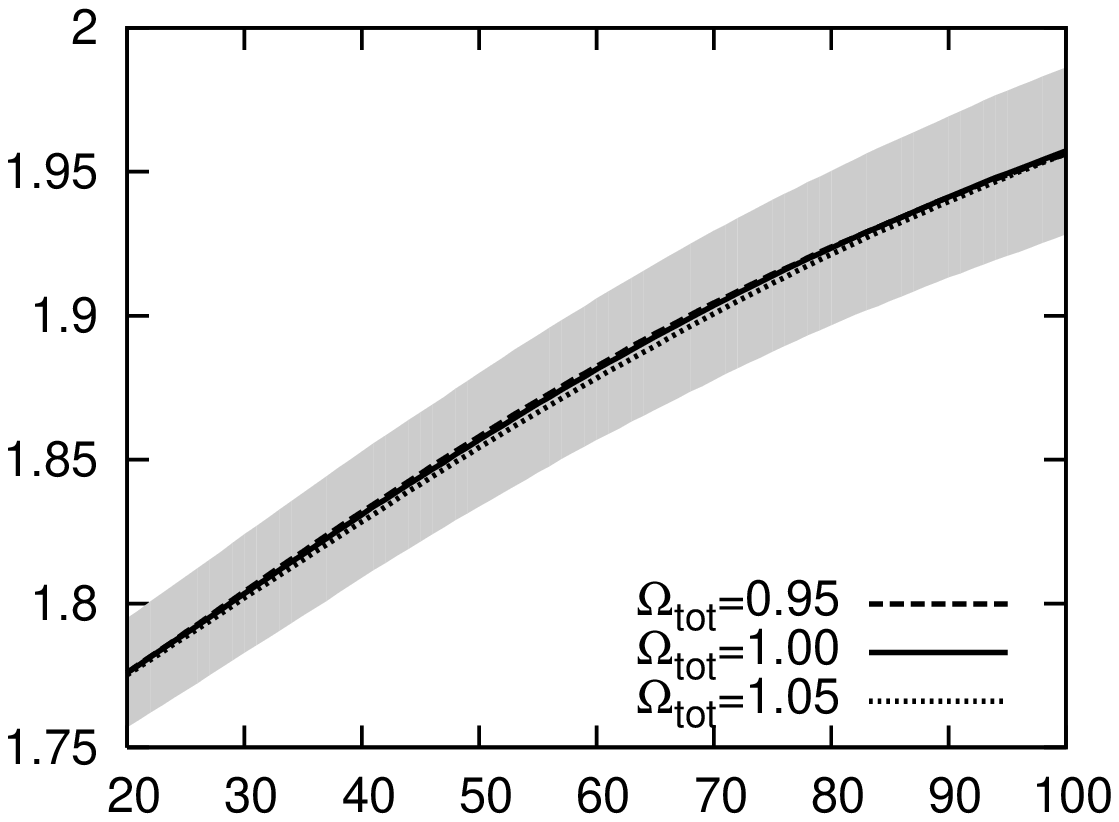}
\includegraphics[angle=-90,width=7.5cm]{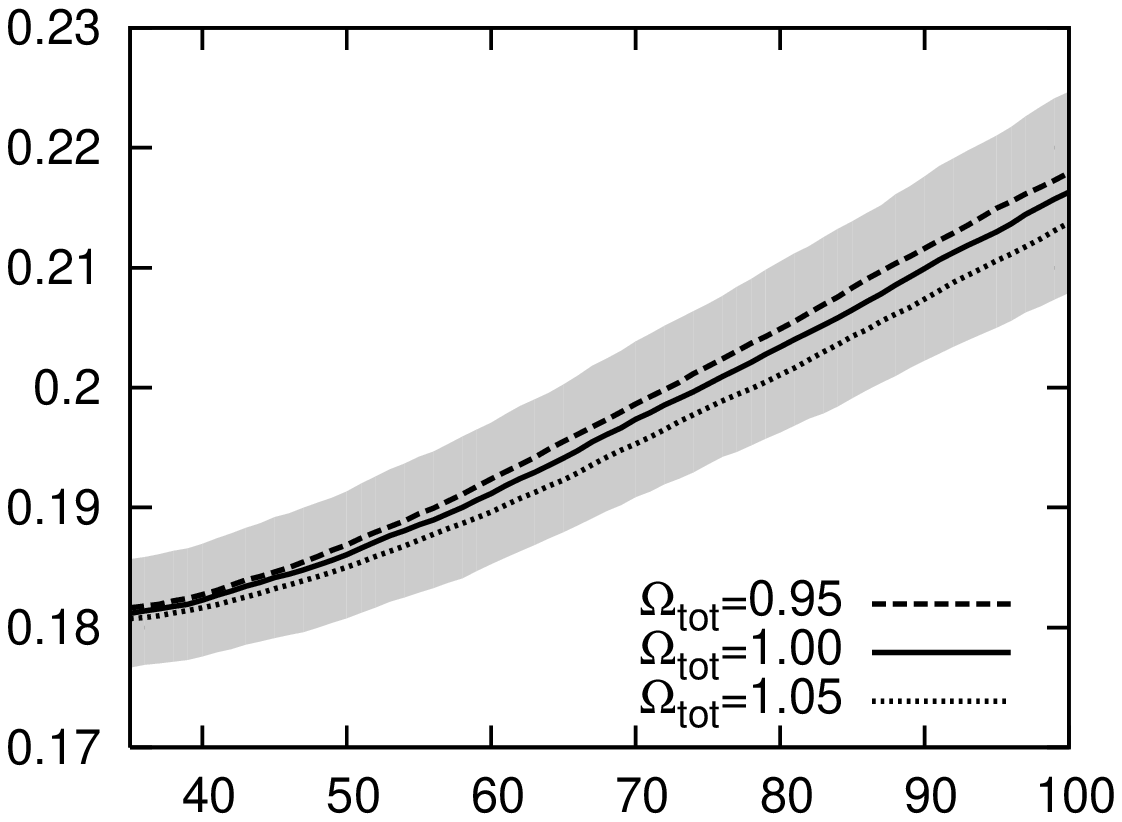}
\end{minipage}
\put(-480,45){ ${\sigma}_{\hbox{\scriptsize fwhm}} = 60\hbox{ arcmin}$}
\put(-260,45){ ${\sigma}_{\hbox{\scriptsize fwhm}} = 60\hbox{ arcmin}$}
\put(-515,52){$\bar{E}$}
\put(-340,-75){$n_{\hbox {\scriptsize min}}$}
\put(-300,55){$\bar{e}$}
\put(-125,-75){$n$}
}
\end{center}
\caption{
\label{fig:versch_fwhm}
The ensemble average of the mean ellipticity
$\bar{E}(\nu, n_{\hbox {\scriptsize min}})$
with $\nu=1$ (left) and the mean elongation $\bar{e}(\nu, n, \Delta n)$
with $\nu=1$ and $\Delta n = 20$ pixels (right)
is computed using three Gaussian smoothing kernels
(from top to bottom: ${\sigma}_{\hbox{\scriptsize fwhm}} = 20\hbox{ arcmin}$,
$40\hbox{ arcmin}$, and $60\hbox{ arcmin}$).
The grey bands reflect the cosmic variance of the flat
$\Lambda$CDM concordance model.
The ensemble average is computed from 1000 realisations.
The sky maps have a resolution of $N_{\hbox{\scriptsize side}}=512$.
}
\end{figure}

In Fig.~\ref{fig:versch_fwhm} the ensemble average of the mean ellipticity
$\bar{E}(\nu, n_{\hbox{\scriptsize min}})$ with $\nu=1$ (left)
and the mean elongation $\bar{e}(\nu,n, \Delta n)$
with $\nu=1$ and $\Delta n = 20$ pixels (right) is shown.
The average is computed from 1000 realisations.
The sky maps possess a resolution of $N_{\hbox{\scriptsize side}}=512$
and $l_{\hbox{\scriptsize max}}=1000$.
The grey band reflects the cosmic variance of the flat
$\Lambda$CDM concordance model (solid line).
This provides the criterion whether two cosmological models can be
distinguished by an elongation measure.
In order to distinguish them the corresponding curves should deviate
more than the width of the band due to the cosmic variance.
Note that in contrast to the previous figures,
the Fig.~\ref{fig:versch_fwhm} displays the elongation as a function
of the spot size and not of the temperature to which the niveau lines belong.
This kind of plotting reveals the remarkable fact
that larger spots possess a higher degree of elongation than smaller ones.
This dependence is not covered by our analytical formulae
which are based on the infinitesimal neighbourhood at maxima or minima.
The impact of three Gaussian smoothing kernels
is also investigated in Fig.~\ref{fig:versch_fwhm} (from top to bottom:
${\sigma}_{\hbox{\scriptsize fwhm}} = 20, 40, 60\hbox{ arcmin}$).
An increased smoothing reduces the degree of elongation.
This is due to the symmetric smoothing kernel
which rounds off the structures.
Here the resolution of $N_{\hbox{\scriptsize side}}=512$ does not resolve
all physical structures of the Gaussian temperature field.
This contrasts to Fig.~\ref{Fig:Ellipticiy_Healpix_fwhm}
which is based on a sky map with a resolution of
$N_{\hbox{\scriptsize side}}=4096$
being sufficiently fine grained to resolve all structures
such that the analytical formulae are applicable. 

In Fig.~\ref{fig:versch_fwhm}
the ensemble average of the elongations is plotted
for three cosmological models
which differ with respect to their curvature. 
In addition to the flat $\Lambda$CDM concordance model,
the ensemble average of a positively curved universe (dotted line)
and a negatively curved universe (dashed line) is displayed.
The figure reveals that a universe with negative curvature possesses
a larger degree of elongation for a given spot size than the flat universe.
This trend is continued for a universe with positive curvature 
which possesses a smaller degree of elongation for a given spot size
than the flat universe.
This encouraging behaviour could be used to distinguish different
cosmological models
if their corresponding curves differ by more than the scattering due
to the cosmic variance.
It turns out that this depends on the resolution of the sky maps.
By applying wider smoothing kernels
(from top to bottom in Fig.~\ref{fig:versch_fwhm})
the differences between the curves decrease and are getting insignificant
compared to the cosmic variance for smoothings around
$\sigma_{\hbox{\scriptsize fwhm}} = 60\hbox{ arcmin}$.
Note that the mean elongation $\bar{e}(\nu, n, \Delta n)$
based on a moving average does a better job than the mean ellipticity
$\bar{E}(\nu, n_{\hbox {\scriptsize min}})$.
This is caused by the fact
that the differences in the elongation vanish for spots with a large area,
and thereby their inclusion blurs the signal.
The figure reveals that a resolution of at least
$\sigma_{\hbox{\scriptsize fwhm}} = 20\hbox{ arcmin}$ is necessary
in order to distinguish between cosmological models having a difference in
curvature of $\Delta \Omega_{\hbox{\scriptsize tot}}=0.05$.
This excludes the application of the ILC sky map of the WMAP team
which has a resolution
of about $\sigma_{\hbox{\scriptsize fwhm}} = 60\hbox{ arcmin}$.
Better resolved sky maps such as the W-band maps of the WMAP team are
available but they are contaminated with noise.
Hence we now focus on the influence of noise on the elongation measures.

\begin{figure}[htb]
\vspace*{-20pt}\begin{center}
\hspace*{-50pt}
{
\begin{minipage}{18.5cm}
\includegraphics[angle=-90,width=7.5cm]{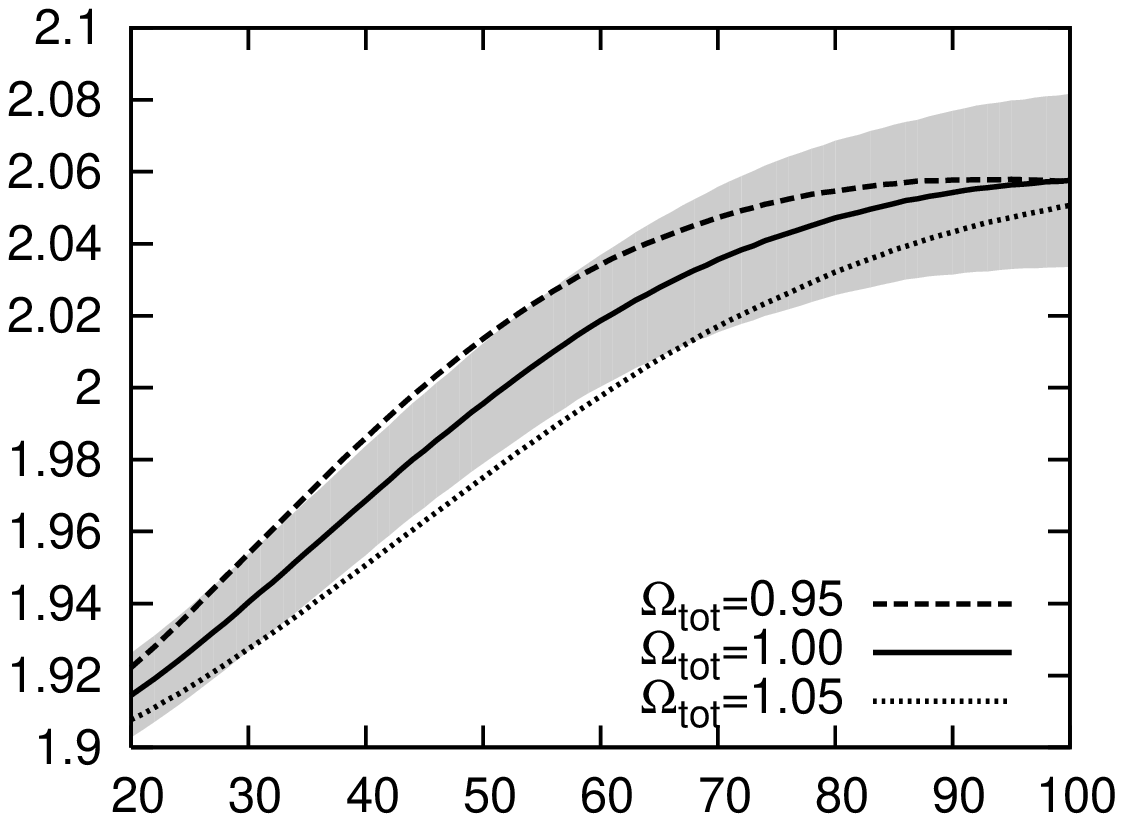}
\includegraphics[angle=-90,width=7.5cm]{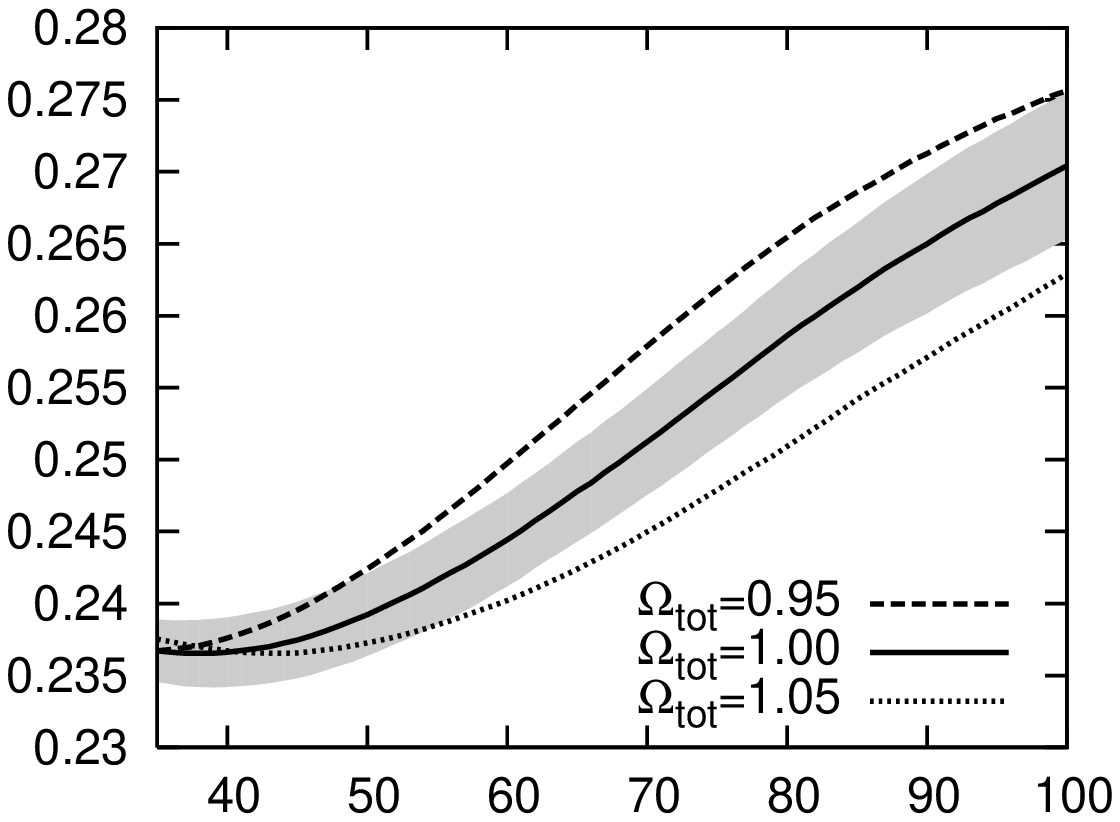}
\end{minipage}
\put(-460,45){without noise}
\put(-240,45){without noise}
\put(-520,52){$\bar{E}$}
\put(-340,-75){$n_{\hbox {\scriptsize min}}$}
\put(-300,55){$\bar{e}$}
\put(-125,-75){$n$}
}
\hspace*{-50pt}
{
\begin{minipage}{18.5cm}
\includegraphics[angle=-90,width=7.5cm]{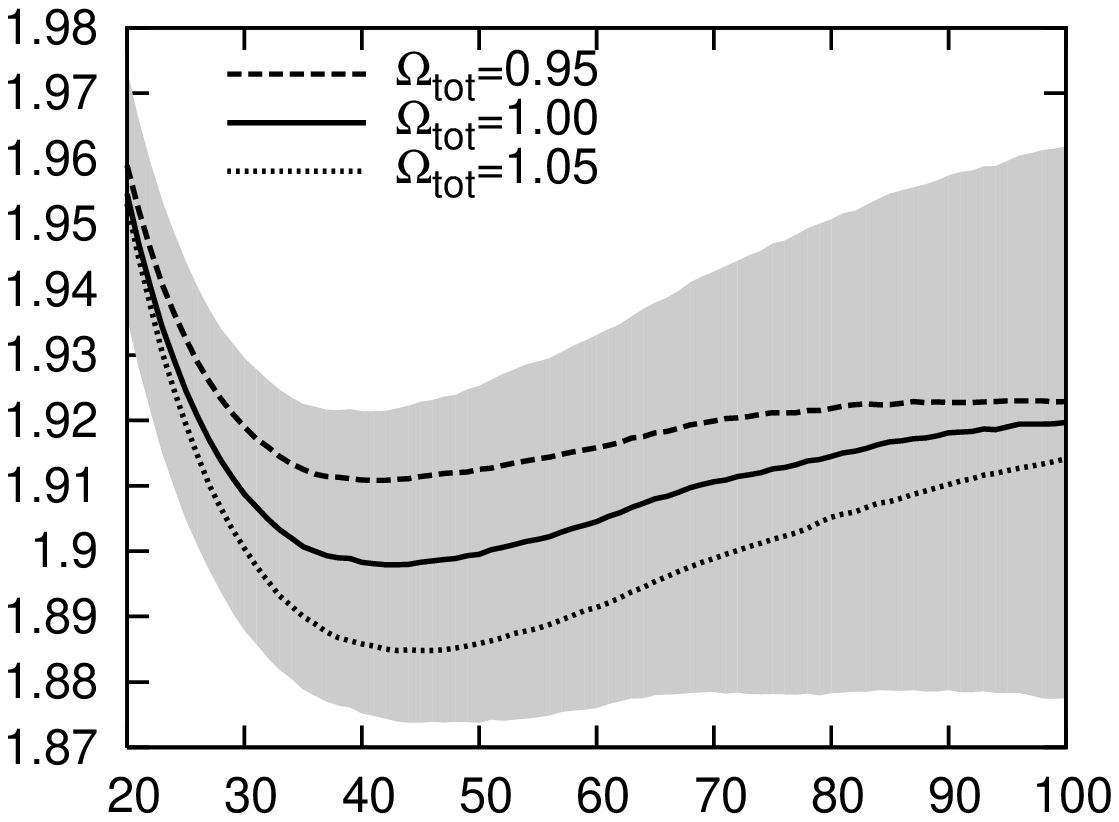}
\includegraphics[angle=-90,width=7.5cm]{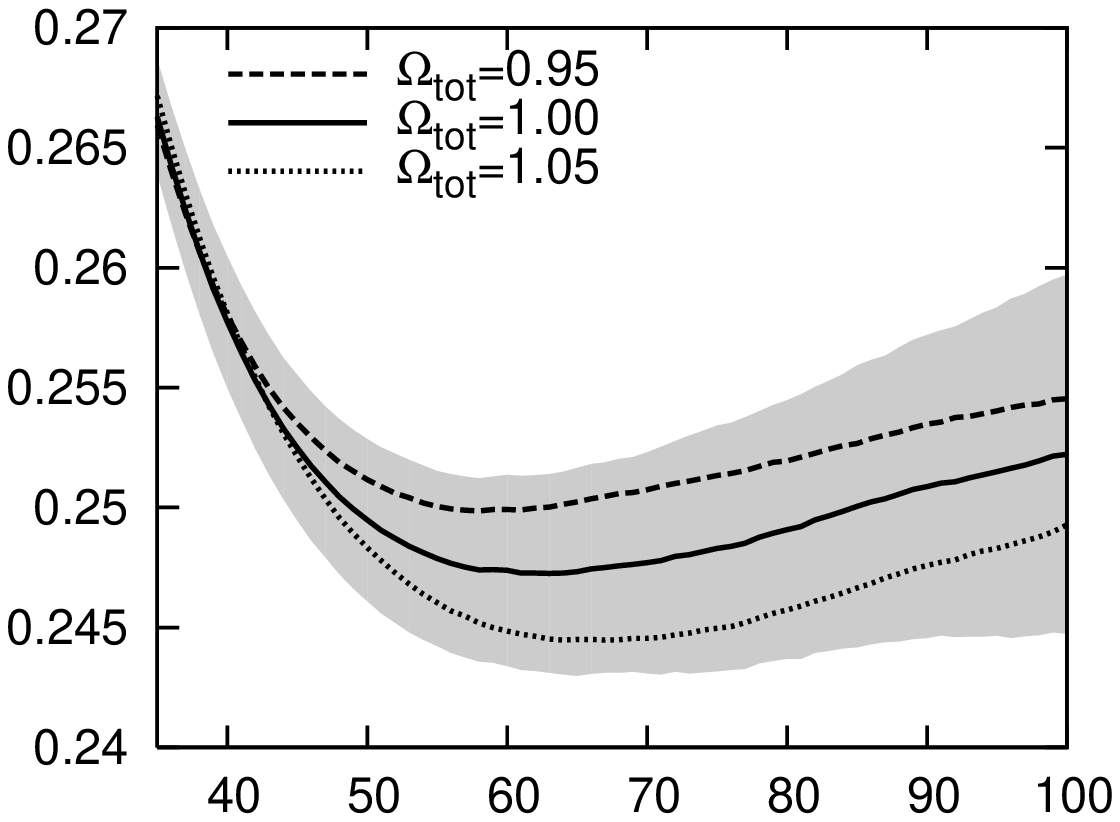}
\end{minipage}
\put(-390,45){with noise}
\put(-170,45){with noise}
\put(-520,52){$\bar{E}$}
\put(-340,-75){$n_{\hbox {\scriptsize min}}$}
\put(-300,55){$\bar{e}$}
\put(-125,-75){$n$}
}
\hspace*{-50pt}
{
\begin{minipage}{18.5cm}
\includegraphics[angle=-90,width=7.5cm]{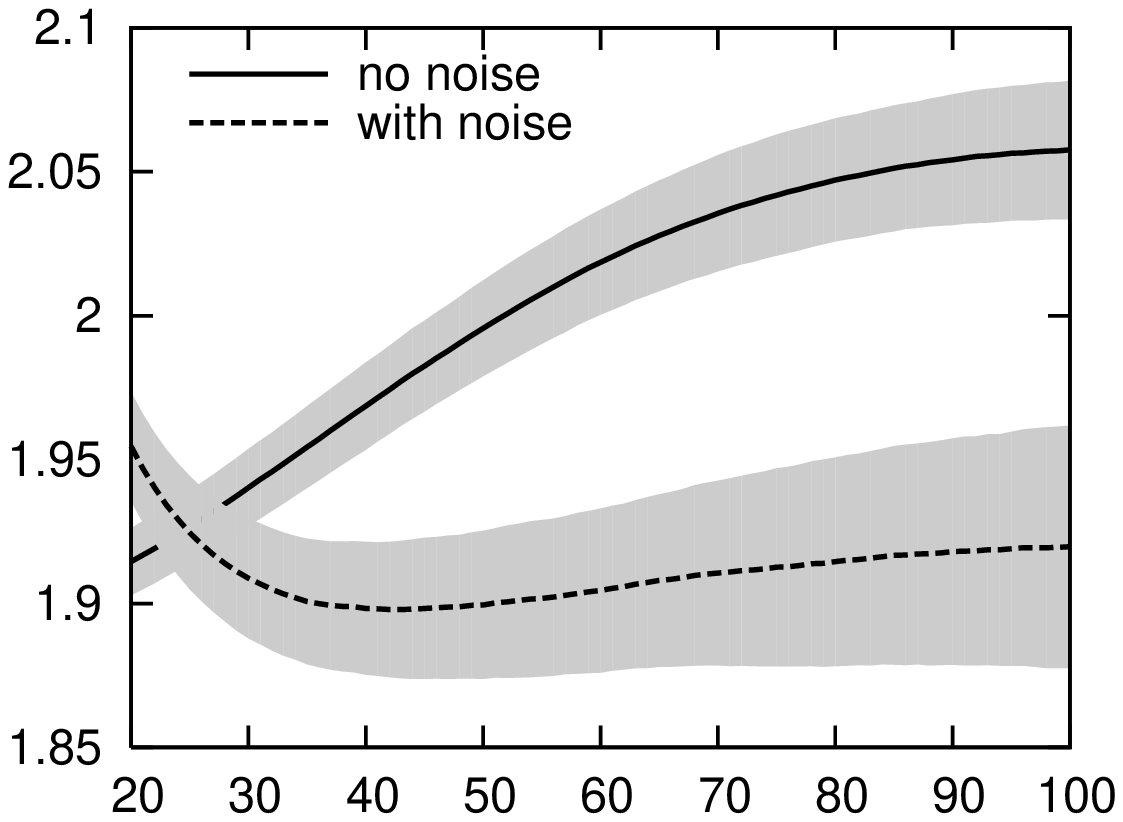}
\includegraphics[angle=-90,width=7.5cm]{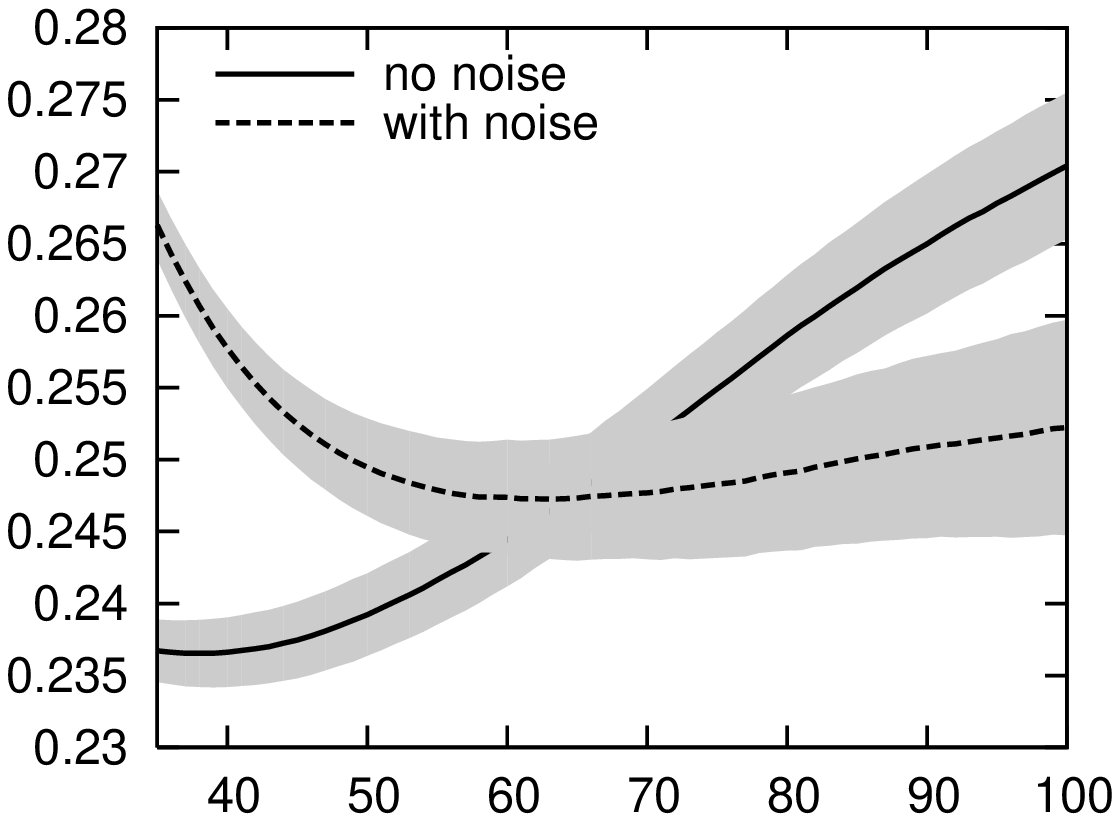}
\end{minipage}
\put(-520,52){$\bar{E}$}
\put(-340,-75){$n_{\hbox {\scriptsize min}}$}
\put(-300,55){$\bar{e}$}
\put(-125,-75){$n$}
}
\end{center}
\caption{
\label{fig:W-Band}
The ensemble average of the mean ellipticity
$\bar{E}(\nu, n_{\hbox {\scriptsize min}})$
with $\nu=1$ (left) and of the mean elongation
$\bar{e}(\nu, n, \Delta n)$ with $\nu=1$ and
$\Delta n = 20$ pixels (right) is presented.
These curves are obtained from 1000 sky simulations which take the
window function of the W-band (channel 4) of WMAP into account.
The grey bands reflect the cosmic variance of the flat $\Lambda$CDM
concordance model (solid curve).
The three rows give the results without noise (top),
with noise (middle) and a comparison of both cases
for the flat $\Lambda$CDM concordance model (bottom).
The sky maps have a resolution of $N_{\hbox{\scriptsize side}}=512$.
}
\vspace*{-10pt}
\end{figure}

In order to address the question whether the W-band maps
are suitable for an elongation analysis,
1000 sky maps are simulated which take into account the
explicit window function of the W-band (channel 4)
of WMAP\cite{Hill_et_al_2009}.
In Fig.~\ref{fig:W-Band} the ensemble averages obtained from such sky maps
are shown for the mean ellipticity
$\bar{E}(\nu, n_{\hbox {\scriptsize min}})$ with $\nu=1$ (left)
and for the mean elongation $\bar{e}(\nu, n, \Delta n)$
with $\nu=1$ and $\Delta n = 20$ pixels (right).
Again the grey band reflects the cosmic variance of the flat
$\Lambda$CDM concordance model (solid line).
In addition, a positively curved universe (dotted line) and
a negatively curved universe (dashed line) is displayed.

The upper row of Fig.~\ref{fig:W-Band} neglects noise completely
and  only takes the W-band window function into account.
Both figures are similar to the two upper figures in
Fig.~\ref{fig:versch_fwhm} which have a resolution of
$\sigma_{\hbox{\scriptsize fwhm}} = 20\hbox{ arcmin}$
comparable to that of the W-band.
The mean elongation $\bar{e}(\nu, n, \Delta n)$
of spots with a small area reveals a saturation
which is absent in Fig.~\ref{fig:versch_fwhm}.
Furthermore, both elongation measures yield smaller values
for spots with a larger area.
This is caused by the application of the W-band window function
which deviates strongly from a Gaussian.
In the middle row of Fig.~\ref{fig:W-Band} the noise is taken into account
which changes the behaviour drastically.
No useful discrimination between the different curved models
is possible any more.
The lowest row in Fig.~\ref{fig:W-Band} shows a comparison
for the $\Lambda$CDM concordance model with and without noise
in one figure in order to emphasise the deteriorating effect of noise.
A further analysis shows
that an incomplete sky coverage additionally broadens the cosmic variance.
Therefore, no comparison to measured data can be shown here.  

\begin{figure}[htb]
\begin{center}
\hspace*{-50pt}\begin{minipage}{18.5cm}
\includegraphics[angle=-90,width=7.5cm]{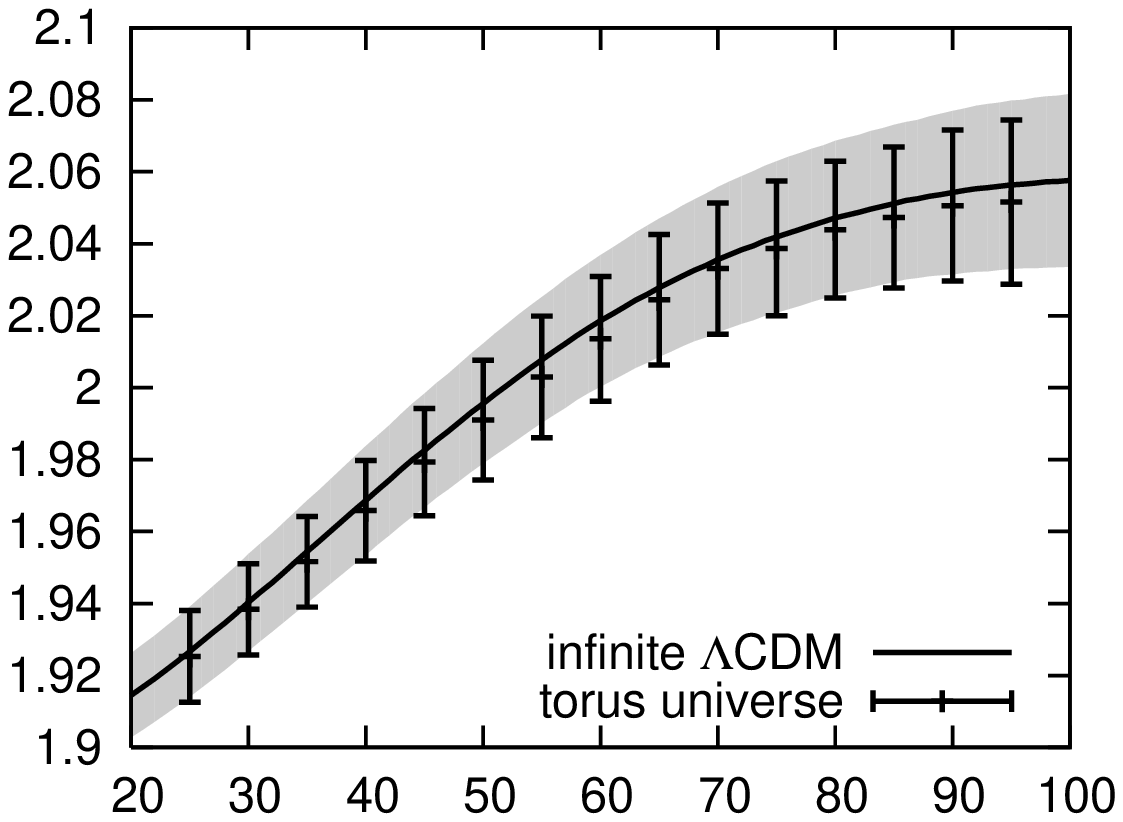}
\includegraphics[angle=-90,width=7.5cm]{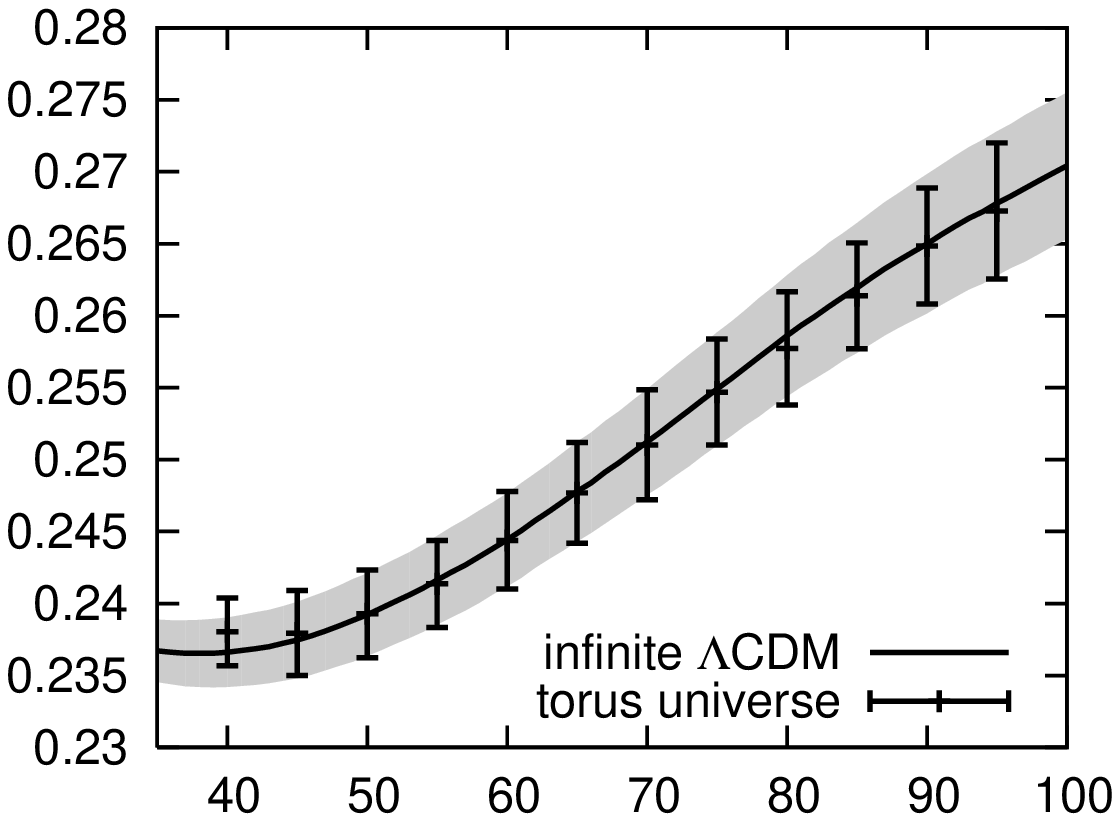}
\end{minipage}
\put(-470,45){(a)}
\put(-235,45){(b)}
\put(-520,52){$\bar{E}$}
\put(-340,-75){$n_{\hbox {\scriptsize min}}$}
\put(-300,55){$\bar{e}$}
\put(-125,-75){$n$}
\end{center}
\caption{\label{Fig:ellipticity_and_elongation_torus}
The ensemble average of the mean ellipticity
$\bar{E}(\nu, n_{\hbox {\scriptsize min}})$
with $\nu=1$ (left) and of the mean elongation
$\bar{e}(\nu, n, \Delta n)$ with $\nu=1$ and
$\Delta n = 20$ pixels (right) is presented.
The elongations $E$ and $e$ of the infinite $\Lambda$CDM model
are shown as a solid curve and its cosmic variance as a grey band
calculated from 1000 realisations.
The elongations of the torus universe with $L=3.86\,L_H$
are shown as error bars which represent the cosmic variance
based on 50 sky simulations.
In both cases the window function of the W-band (channel 4) of WMAP 
is taken into account, but the noise is neglected.
The sky maps have a resolution of $N_{\hbox{\scriptsize side}}=512$.
}
\end{figure}

Finally, we compare the infinite volume $\Lambda$CDM model
with a statistically anisotropic multi-connected model
for which we again choose the cubic topology with side length
$L=3.86\,L_H$.
As already discussed in Section \ref{Number_of_Maxima_and_Minima}
the volume of the fundamental cell is a significant fraction
of the volume inside the surface of last scattering.
Therefore, only the largest angular scales are modified
but not the fine structures which determine the elongation.
In Fig.~\ref{Fig:ellipticity_and_elongation_torus}
the mean ellipticity $\bar{E}$ and the elongation $\bar{e}$
computed from 50 torus realisations are compared with
the elongations of the infinite flat $\Lambda$CDM model.  
As expected the mean values are almost identical
and their small differences are confined within the
corresponding bands of cosmic variance.
Thus, the anisotropy of the torus model is too weak
to allow a discrimination between the models
using the elongation measures.

In addition to the above investigations
we also analysed our simulated sky maps at other thresholds $\nu$.
Quantitatively our statements are equivalent for hot spots and cold spots
which is due to the isotropy properties of the random field.
So we restrict ourselves in this paper to hot spots only.
Nevertheless, if a single realisation of a measured sky map is studied,
both should be combined into one statistic for a better significance.
Other thresholds $\nu$ with $\nu\neq\pm 1$ possess a smaller number of spots
and, therefore, lead to larger variances
which is counterproductive in order to distinguish between cosmological models.
Besides the mean values $\bar{E}(\nu, n_{\hbox {\scriptsize min}})$ and
$\bar{e}(\nu, n, \Delta n)$
we also studied higher moments, higher central moments and
their statistical interpretations.
Since it turns out that the mean value provides the most robust measure
we restrict our discussion to them here.
The moving average applied in $\bar{e}(\nu, n, \Delta n)$ leads
to the best discrimination between cosmological models,
but the additional parameter $\Delta n$ has to be chosen adequately,
since too small values result in small spot numbers (lack on statistics)
and too large values result in a blurring caused by large spots.


\section{Summary}


In this paper the structures of CMB sky maps are studied
with respect to the crucial question
whether these structures betray some information
about the underlying cosmological model.
There are various quantities to analyse the structures of the niveau lines
and the focus is put on the elongation
which can be described by the ellipticity $E$, Eq.~(\ref{Eq:def_ellip}),
or the elongation measure $e$ defined in Eq.~(\ref{Eq:def_e}).
The elongation depends on the parameter $\alpha$, Eq.~(\ref{Eq:alpha}),
which in turn depends on the multipole spectrum $C_l$
and thus on the cosmology.
The theoretical dependence of various elongation measures on $\alpha$
is given for the case that the statistical properties of
the temperature field are those of two-dimensional homogeneous and
isotropic Gaussian random fields.
Our analysis leads to the conclusion
that this dependence, however, is weak.
Since the resolution of the maps has a superior influence on the niveau lines,
the results obtained from sky maps depend on the pixelization
and on the beam profile of the detector.
Both deteriorating restrictions are analysed.
It is found that the elongation cannot be analysed without reference
to the pixelization and the beam profile.

The dependence of the elongation on the cosmological parameters can be
investigated for niveau lines classified either by their temperature
or by the area which they encompass.
The curvature of the universe is revealed by the elongation
as demonstrated in Fig.~\ref{fig:versch_fwhm},
where the elongation is considered as a function of the size of the spots.
The best discrimination between cosmological models is obtained
by applying the moving average to $\bar{e}(\nu, n, \Delta n)$.
A resolution of at least $\sigma_{\hbox{\scriptsize fwhm}} = 20\hbox{ arcmin}$
is required in order to distinguish between cosmological models
which differ in the curvature by $\Delta \Omega_{\hbox{\scriptsize tot}}=0.05$.
This excludes the analysis of the ILC map of the WMAP team,
whose resolution is three times lower.
The required resolution is achieved by the W-band sky map
which has, however, much stronger noise.
The analysis of simulated maps with the noise properties of the  W-band map
shows that it is also not suitable for this investigation.
The Planck mission will provide maps with a significantly higher
resolution as well as lower noise
such that there is the hope that the results of that mission
can be analysed with respect to the structure properties of the CMB.

We also study the elongation properties of a multi-connected space form,
i.\,e.\ a model of the universe having a non-trivial topology,
because it is not statistically isotropic and thus violates the assumptions
which have to be satisfied for the analytical expressions to be valid.
No distinction is found between the trivial topology,
i.\,e.\ the concordance model, and the non-trivial topology
with respect to the elongation.
This is expected since the topology modifies the physics on the largest
scales and has thus only a modest influence on the elongation.
At least as long as the fundamental cell is so large
that it constitutes a significant fraction within the
surface of last scattering.
Thus the best prospects for elongation measures can be found
in the detection of curvature.

\appendix

\section[The  Gaussian Random Field and the CMB]
{The  Gaussian Random Field and the CMB}
\label{appendix_gaussian_random_field}

In this Appendix, we outline the derivation of the peak density and
the ellipticity of  the temperature fluctuations $\delta
T\left(\hat{n} \right)$ of the CMB under the assumption that the
temperature fluctuations behave as a homogeneous and isotropic Gaussian
random field
\begin{equation}
\label{Gauss_deltaT}
f\left(\delta T\left(\hat{n} \right) \right)= \frac{1}{\sqrt{2 \pi
\sigma_0^2 }}
   \exp \left(-\frac{\left|\delta T\left(\hat{n} \right) \right|^2}{2
\sigma_0^2 }\right)
\hspace{10pt}
\end{equation}
on the 2-dimensional sphere.
Here $\sigma_0^2 \left(\hat{n} \right):=\left \langle \delta T
\left(\hat{n} \right)\delta T \left(\hat{n} \right) \right \rangle$ is
the variance of the temperature fluctuations. 
In an isotropic model $\sigma_0^2$ is independent of the direction $\hat{n}$
and  is given in that case by 
\begin{equation}
\label{sigma_0}
\sigma_0^2 = \sum_{l}\frac{\left(2l+1\right)}{4\pi} C_l
\left|F_{l}\right|^{2}  = C(0)
\hspace{10pt}.
\end{equation}
The ensemble average is specified as $\langle ...  \rangle$, 
$F_{l}$ accounts for the appearance of a beam profile 
and $C(0)$ is the correlation function of the temperature fluctuations 
in a homogeneous and isotropic universe at a separation angle $\vartheta=0$.
To use a compact notation in this Appendix, 
the subscript $F_{l}$ at the temperature field $\delta T_{F_l}$
is omitted. 
This index stands for the smoothing operation (\ref{Eq:smooth1}).
The following derivations and notations are based 
on Ref.~\refcite{Bond_Efstathiou_1987}.

The  temperature fluctuation $\delta T\left(\hat{n}\right)$ close to 
the direction $\hat{n}'$ on the 2-sphere is given by the Taylor series
\begin{eqnarray} \nonumber
\delta T\left(\hat{n}\right)  & = &
\delta T\left(\hat{n}'\right) \; + \;
\left[\nabla_i \delta T\left(\hat{n}\right)\right]_{\hat{n}=\hat{n}'}
(x^{i} - x'^i)
\\ & &
\label{Taylorentwicklung_dT}
\hspace{10pt} + \;
\frac{\left[\nabla_i \nabla_j \delta T \left(\hat{n} \right)
\right]_{\hat{n}=\hat{n}'}}{2} (x^{i} - x'^i)(x^{j} - x'^j)
\; + \; \dots
\hspace{10pt}.
\end{eqnarray}
$\nabla_i$ is the covariant derivative on the 2-sphere with respect to
the coordinate $x^i$.
In the sequel we will choose $x^1=\theta$ and $x^2=\phi$.
The corresponding line element is given by $ds^2 = d\theta ^2 + \sin^2
\theta d \phi ^2$.
The associated covariant metric is $\gamma_{11} = 1$, $\gamma_{22} = \sin^2 \theta$
 and $\gamma_{ij}= 0$ otherwise.
Therefore the Christoffel symbols are
$\Gamma^{2}_{12} = \Gamma^{2}_{21} = \frac{\cos \theta}{\sin \theta}$,
$\Gamma^{1}_{22} = -\sin \theta \cos \theta$ and
$\Gamma^{k}_{ij} = 0$ otherwise.

The  Taylor series~(\ref{Taylorentwicklung_dT}) can be written in a compact way by introducing
the variables
$\delta T=\delta T\left(\hat{n}\right)$,
$\eta_i=\eta_i\left(\hat{n}\right) :=  \nabla_i \delta T\left(\hat{n}\right)$ and 
$\zeta_{ij}=\zeta_{ij}\left(\hat{n}\right)
:= \nabla_i \nabla_j \delta T\left(\hat{n}\right)$.
Calculating the symmetric correlation matrix for Gaussian random fields
with respect to the ensemble average in a homogeneous and isotropic model,
one obtains
\begin{eqnarray} \nonumber
& &\left \langle \delta T \delta T \right \rangle = \sigma_{0}^{2}, \;
\; \; \left \langle \delta T \eta_i \right \rangle = 0 , \; \; \;
\left \langle \eta_i \eta_j \right \rangle =  \frac{\sigma_{1}^{2}}{2}
\gamma_{ij} , \; \; \; \left \langle \delta T \zeta_{ij} \right
\rangle =  -\frac{\sigma_{1}^{2}}{2} \gamma_{ij} ,
\\ & &
\label{Korrelation_dT_nabladT_nabalanabladT}
\left \langle \eta_i \zeta_{js} \right \rangle = 0, \; \; \; \left
\langle \zeta_{ij} \zeta_{rs} \right \rangle =
\frac{\sigma_{2}^{2}}{8}\left[\gamma_{ij}\gamma_{rs}+\gamma_{is}
\gamma_{rj}+\gamma_{js}\gamma_{ir}\right]+\frac{\sigma_{1}^{2}}
{2}\gamma_{ij}\gamma_{rs}
\end{eqnarray}
with
\begin{equation}
\label{sigma_1}
\sigma_1^2:=\sum_{l}\frac{\left(2l+1\right)}{4\pi} C_l  \left|F_{l}\right|^{2} l(l+1)
\hspace{10pt}
\end{equation}
and
\begin{equation}
\label{sigma_2}
\sigma_2^2:=\sum_{l}\frac{\left(2l+1\right)}{4\pi} C_l  
\left|F_{l}\right|^{2} (l-1)l(l+1)(l+2)
\hspace{10pt}.
\end{equation}

\subsection[The peak density of the CMB]{The peak density of the CMB}
\label{appendix_peak_density}

Now the formulae for the number of extrema per solid angle
are derived from the density of extrema $n_{\hbox{\scriptsize ext}}(\hat{n})=
\sum_{p} \delta \left(\hat{n}-\hat{n}_{p} \right)$ for
an isotropic and homogeneous Gaussian random temperature field
$\delta T\left(\hat{n}\right)$  on the 2-sphere.
The density of extrema turns out to be given by
\begin{equation}
\label{Peakdensity}
n_{\hbox{\scriptsize ext}}(\hat{n})= \delta \left(\vec{\eta}(\hat{n})\right)
\; \left| \det(\zeta(\hat{n}))\right|
\hspace{10pt}
\end{equation}
by using $x^{i} - {x^{i}}_p \approx
\eta_j\left(\hat{n}\right)\left(\zeta^{ji}\right)^{-1}(\hat{n}_p)$.
Here $\hat{n}_p$ denotes the directions of the extrema.

We choose $\theta=\frac{\pi}{2}$ without loss of generality.
We transform $\zeta_{ij}$ onto its principal coordinate system 
by a rotation with an angle $\tilde{\theta}$,
thus obtaining the diagonal form $-\hbox{diag}(\lambda_1,\lambda_2)$,
ordered by $\left|\lambda_1\right| \ge \left|\lambda_2\right|$.
At maxima or minima of the temperature field the eigenvalues 
$\lambda_1$ and $\lambda_2$ have both positive or both negative values,
respectively.
At saddle points one of these eigenvalues is positive and
the other is negative. 
In the following only maxima and minima are considered. 
The eigenvalues of the Hessian contain slightly different information
as the eigenvalues of the inertia tensor of
the area enclosed by the contour lines which are used to
calculate the ellipticity $E$ in Section \ref{Elongation_HS_CS}.
Now with the eigenvalues $\lambda_1$ and $\lambda_2$, 
the following variables are defined 
\begin{equation}
\label{xe_lambda1_lambda2}
x:=\frac{\lambda_1+\lambda_2}{\sigma_2\,\alpha}\hspace{5pt}, \hspace{1cm}
e:=\frac{\lambda_1-\lambda_2}{2\,(\lambda_1+\lambda_2)} 
  = \frac{\lambda_1-\lambda_2}{2\,\sigma_2\,\alpha\,x}
\hspace{5pt},
\end{equation}
in terms of which the Hessian $\zeta_{ij}$ reads
\begin{eqnarray}
\label{zetas}
\zeta_{11} & = &
-\frac{\sigma_2\,x\,\alpha}{2}\Big[1+2\,e\,\cos(2\tilde{\theta})\Big]\hspace{5pt},
\\
\zeta_{22} & = &
-\frac{\sigma_2\,x\,\alpha}{2}\Big[1-2\,e\,\cos(2\tilde{\theta})\Big]\hspace{5pt},
\\
\zeta_{12} & = & -\sigma_2\,x\,\alpha\,e\,\sin(2\tilde{\theta})\hspace{5pt},
\hspace{10pt}
\end{eqnarray}
where $\alpha:=\sqrt{1+\frac{2\,\sigma_1^2}{\sigma_2^2}}$ and 
$e$ is the elongation, Eq.~(\ref{Eq:def_e}).
It should be noted that in Ref.~\refcite{Bond_Efstathiou_1987} 
$e$ is termed ellipticity. 
In general these new variables are restricted to 
$\tilde{\theta} \in [0,\pi]$, $x \in (-\infty,\infty)$ and $e \in [0,\infty)$. 
In case of extrema the interval of the elongation is confined to
$e \in [0,\frac{1}{2})$. 

The transformation of the volume element in $\zeta$-space is
\begin{equation}
\label{volume_zeta}
d\zeta_{11}d\zeta_{22}d\zeta_{12} \; = \;
2\,\sigma_2^3\,\alpha^3\,x^2\,dx\,de\,d\tilde{\theta}
\hspace{5pt} .
\end{equation}
The probability distribution for the variables
$\nu:=\frac{\delta T}{\sigma_0}$,
$\vec{\eta}$, $x$, $e$, and $\tilde{\theta}$ is given by
\begin{eqnarray} \nonumber
P(\nu,\vec{\eta},x,e,\tilde{\theta})\,de\,dw\,dx\,d\tilde{\theta}\,
d^{2}\vec{\eta}
& = & \exp\left[-\frac{w^2}{2}\right] \exp
\left[-\left(\frac{1}{2}+4e^{2}\alpha^{2}\right)x^2\right]
\\ & & \hspace{-75pt}
\label{verteilung_x_eta_e_theta}
\times \; \exp \left[-\frac{\vec{\eta}^{\,2}}{\sigma^2_1} \right] \,
8\,\left(\alpha x\right)^{2}\,e\,de\,\frac{dw}{\sqrt{2\,\pi}} \,
\frac{dx}{\sqrt{2\,\pi}}\,\frac{d\tilde{\theta}}{\pi}
\frac{d^{2}\vec{\eta}}{\pi\sigma_1^2}
\hspace{10pt} 
\end{eqnarray}
where the result is simplified by using the variable
$w:=\frac{\nu-\gamma x}{\sqrt{1-\gamma^2}}$ and introducing the abbreviation
$\gamma:=\frac{\sigma_1^2}{\sigma_2\,\sigma_0\,\alpha}$. 
$\gamma$ is determined by the power spectrum.
Here $w$ and $x$ are independent and normalised
$\langle x^2 \rangle=1$, $\langle w^2 \rangle=1$.

As a result of the restriction to positive eigenvalues
$\lambda_1 \ge \lambda_2 \ge 0$,
i.e. restricting to maxima, one obtains $x \in [0,\infty)$ 
and $e \in [0,\frac{1}{2})$.
Using this, 
$\theta^{*2}:=\frac{2\,\sigma_1^2}{\sigma_2^2} = \alpha^2 -1$ 
and
\begin{equation}
\label{det_zeta}
\det(\zeta)=\frac{1}{4}\,\sigma_2^2\,x^2\,\alpha^2\,(1-4\,e^2)
\hspace{10pt},
\end{equation}
we obtain for the mean differential density of maxima
\begin{eqnarray} \nonumber
\label{N_max_exvutheta}
N_{\hbox{\scriptsize max}}(\nu,x,e,\tilde{\theta}) \; &=& \;
\int_{-\infty}^{\infty}\,dw'\int_{\mathbb{R}^2}\,d^{2}\vec{\eta}\,'
\int_{0}^{\infty} dx' \int_{0}^{\frac{1}{2}} de'
\int_{0}^{\pi} d\tilde{\theta}\,'
\;P(\nu',\vec{\eta}\,',x',e',\tilde{\theta}')
\\ & & \nonumber \hspace{20pt} \times
\delta(\vec{\eta}\,')\,\delta(x'-x)\,\delta(e'-e)\,
\delta(\tilde{\theta}\,'-\tilde{\theta})\,
\frac{1}{4}\,\sigma_2^2\,x'^2\,\alpha^2\, (1-4\,e'^2)
\\ & &\nonumber
=\frac{2}{\pi^3\,\theta ^{*2}\,\sqrt{1-\gamma^2}}
e\,\left(1-4e^2\right)\left(\alpha x\right)^{4}
\exp\left[-\frac{w^2}{2}\right]
\\ & & \hspace{20pt} \times
\exp \left[-\left(\frac{1}{2}+4e^{2}\alpha^{2}\right)x^2\right]
\hspace{5pt}.
\end{eqnarray}
The integration over the orientation angle $\tilde{\theta}$ yields
($w=w(x,\nu)$)
\begin{eqnarray} \nonumber
N_{\hbox{\scriptsize max}}(\nu,x,e) & = &
\frac{2}{\pi^2\,\theta ^{*2}\,\sqrt{1-\gamma^2}}
e\,\left(1-4e^2\right)\left(\alpha x\right)^{4}
\exp\left[-\frac{w^2}{2}\right]
\\ & & \hspace{80pt}
\label{N_max_exvu}
\times \; \exp \left[-\left(\frac{1}{2}+4e^{2}\alpha^{2}\right)x^2\right]
\hspace{5pt}.
\end{eqnarray}
An integration over $\nu$ in the last expression results in
\begin{equation}
\label{N_max_ex}
N_{\hbox{\scriptsize max}}(x,e)=
\frac{1}{\theta^{*2}}
\left(\frac{2}{\pi}\right)^{\frac{3}{2}}\left(\alpha x\right)^{4}
e\left(1-4e^2\right)\exp\left[-\left(\frac{1}{2}+4e^2\alpha^2\right)x^{2}\right]
\hspace{5pt}
\end{equation}
where we have substituted the variable $\nu$ by $w$.
A further integration over $x$ yields
\begin{equation}
\label{N_max_e}
N_{\hbox{\scriptsize max}}(e)=\frac{6\,e\left(1-4e^2\right)\alpha^{4}}
{\theta^{*2}\pi\left(1+8\alpha^2e^2\right)^{\frac{5}{2}}}
\hspace{5pt}.
\end{equation}
Here we have used the substitution
$y=\left(\frac{1}{2}+4e^2\alpha^2\right)x^{2}$.
A final integration over $e$ leads to the total number of maxima
per solid angle
\begin{eqnarray} \nonumber
N_{\hbox{\scriptsize max}}& = &
\int_{0}^{\frac{1}{2}} de\, N_{\hbox{\scriptsize max}}(e)
\; = \;
\frac{3a^4}{8\pi\theta^{*2}}\phantom{i}_{2}F_1(1,\frac{5}{2};3;-2\alpha^2)
\\ & = &  \nonumber
\label{N_max}
\frac{\left(\alpha^{2}-1\right)\sqrt{1+2\alpha^{2}}+1}
{4\pi\theta^{*2}\sqrt{1+2\alpha^{2}}}
\\ & = &
N_{\hbox{\scriptsize max}}^{\hbox{\scriptsize B\&E}}\,
\frac{\left[\left(\alpha^{2}-1\right)\sqrt{1+2\alpha^{2}}+1\right]\sqrt{3}}
{\sqrt{1+2\alpha^{2}}}
\hspace{5pt}
\end{eqnarray}
where $N_{\hbox{\scriptsize max}}^{\hbox{\scriptsize B\&E}}:=
\frac{1}{4\pi\theta^{*2}\sqrt{3}}$
is the limiting case derived in Ref.~\refcite{Bond_Efstathiou_1987}.
We achieved the first representation of $N_{\hbox{\scriptsize max}}$ by using 
the substitution $y=e^{2}$ and the integral~(2.2.6.15) in 
Ref.~\refcite{Prudnikov_Brychkov_Marichev_1988_Band1}
for  the hypergeometric function $\,_2F_1(a,b;c;z)$.

Integrating Eq.~(\ref{N_max_exvu}) with respect to $x$ and
using Eq.~(2.3.15.3) 
in Ref.~\refcite{Prudnikov_Brychkov_Marichev_1988_Band1} or Eq.~(2.3.15.7) in
Ref.~\refcite{Prudnikov_Brychkov_Marichev_1988_Band1},
we obtain two representations of 
$N_{\hbox{\scriptsize max}}(\nu,e)$,
\begin{eqnarray} \nonumber
\label{N_max_evu}
N_{\hbox{\scriptsize max}}(\nu,e) & = &
\frac{48 \left(1-\gamma^2\right)^2\, \alpha^4 \,e\left(1-4e^2\right)}
{\pi^2\,\theta ^{*2} \left(8e^2\alpha^2
\left(1-\gamma^2\right)+1\right)^{\frac{5}{2}}}
\\ & & \nonumber \hspace{50pt} \times
\exp\left[-\nu^{2}\, \frac{2\left(8e^2\alpha^2
\left(1-\gamma^2\right)+1\right)-\gamma^2}{4\left(1-\gamma^2\right)
\left(8e^2\alpha^2 \left(1-\gamma^2\right)+1\right)}\right]
\\ & & \nonumber \hspace{50pt} \times
D_{-5} \left(\frac{-\gamma
\nu}{\sqrt{\left(1-\gamma^2\right)\left(8e^2\alpha^2
\left(1-\gamma^2\right)+1\right)}}\right)
\\ & = & 
\frac{\sqrt{2\pi}e\left(1-4e^2\right)\alpha^4}{\pi^2\,\theta^{*2}\sqrt{8e^2\alpha^2
\left(1-\gamma^2\right)+1}}
\exp\left[-\frac{\nu^{2}}{2\left(1-\gamma^{2}\right)}\right]
\\ & & \nonumber \hspace{50pt} \times
\frac{\partial^{4}}{\partial q^{4}} \left[\exp\left[\frac{q^2}{4p}\right]
\hbox{erfc}\left(\frac{q}{2\sqrt{p}}\right)\right]
\hspace{10pt}
\end{eqnarray}
with $q=-\frac{\gamma \nu}{\sqrt{1-\gamma^2}}$ and $p=\frac{8e^2\alpha^2
\left(1-\gamma^2\right)+1}{2\left(1-\gamma^2\right)}$.
$D_{\alpha}(x)$ is the parabolic cylinder function and
$\hbox{erfc}(x)$ the complementary error function.

Integrating Eq.~(\ref{N_max_exvu}) over $e$, one gets
\begin{equation}
\label{N_max_xvu}
N_{\hbox{\scriptsize max}}(\nu,x) = \frac{1}{4\pi^2\,\theta
^{*2}\sqrt{1-\gamma^2}}
\exp\left[-\frac{w^2}{2}\right] \exp \left[-\frac{x^2}{2}\right] f(x)
\hspace{10pt}
\end{equation}
with $f(x):=
\exp\left[-\left(\alpha x\right)^{2}\right]-1+\left(\alpha x\right)^{2}$.
An integration of Eq.~(\ref{N_max_xvu}) over $\nu$ or $x$ yields
\begin{equation}
\label{N_max_x}
N_{\hbox{\scriptsize max}}(x) =
\frac{1}{(2\pi)^{\frac{3}{2}}\,\theta ^{*2}}
   \exp \left[-\frac{x^2}{2}\right] f(x)
\hspace{10pt}
\end{equation}
and
\begin{equation}
\label{N_max_nu}
N_{\hbox{\scriptsize max}}(\nu) = \frac{1}{(2\pi)^\frac{3}{2}\,\theta^{*2}}
\exp\left[-\frac{\nu^2}{2}\right] \,G(\nu,\gamma,\alpha)
\hspace{5pt}
\end{equation}
with
\begin{eqnarray} \nonumber
G(\nu,\gamma,\alpha)&:=&\gamma \nu (1-\gamma^2) \frac{\exp\left[-\frac{\gamma^2\,\nu^2}
{2(1-\gamma^2)}\right]}{\sqrt{2\,\pi\,(1-\gamma^2)}}
\\ & & \hspace{0pt}\nonumber
+\left(\alpha^2(1-\gamma^2)-1+\gamma^2\nu^2\right)
\left[1-\frac{1}{2}\hbox{erfc}
\left( \frac{\gamma  
\nu}{\sqrt{2(1-\gamma^2)}}\right) \right]
\\ & & \hspace{0pt}
\nonumber
+\frac{\exp\left[\frac{-\alpha^2\,\gamma^2\,\nu^2}
{1+2\alpha^2(1-\gamma^2)}\right]}{\sqrt{2\alpha^2(1-\gamma^2)+1}}
\left[1-\frac{1}{2}\hbox{erfc}
\left( \frac{\gamma  
\nu}{\sqrt{2(1-\gamma^2)(1+2\alpha^2(1-\gamma^2))}}\right)\right]
\hspace{5pt},
\end{eqnarray}
respectively.
We have used the integral~(2.3.15.7) in
Ref.~\refcite{Prudnikov_Brychkov_Marichev_1988_Band1}
and the relation $\hbox{erfc}(-x)=2-\hbox{erfc}(x)$ in order to compute 
$N_{\hbox{\scriptsize max}}(\nu)$.

The mean differential density of maxima
depending on the eccentricity
$\varepsilon:=\sqrt{1-\lambda_2/\lambda_1}\equiv 2[e/(1+2e)]^{1/2}$ 
($\varepsilon \in [0,1)$) is given by
\begin{equation}
\label{N_max_epsilon}
N_{\hbox{\scriptsize max}}(\varepsilon) \; = \;
\frac{3\,\sqrt{2}\, \varepsilon^3(1-\varepsilon^2)}
{\pi \, \theta ^{*2} \,\alpha \left(\frac{(2-\varepsilon^2)^2}{2\alpha^2} +
\varepsilon^4\right)^{\frac{5}{2}}}
\hspace{5pt}
\end{equation}
and depending on the ellipticity
$E:=\sqrt{\lambda_1/\lambda_2}\equiv [(1+2e)/(1-2e)]^{1/2}$
($E \in [1,\infty)$) by
\begin{equation}
\label{N_max_gE}
N_{\hbox{\scriptsize max}}(E) = \frac{24\,\alpha^4\, E^3(E^2-1)}
{\pi \, \theta ^{*2} [E^4(1+2\alpha^2)+2\,E^2\,(1-2\alpha^2)+(1+2\alpha^2)]^{\frac{5}{2}}}
\hspace{5pt}.
\end{equation}

We obtain the corresponding expressions for the mean differential densities 
of the minima from those of 
the maxima by replacing $\nu$ by $-\nu$ and $x$ by $-x$, e.g.
$N_{\hbox{\scriptsize min}}(\nu)=N_{\hbox{\scriptsize max}}(-\nu)$.
Using this we get the combined distribution
\begin{eqnarray}
\label{N_min_nu_N_max_nu}
\left(N_{\hbox{\scriptsize max}}+N_{\hbox{\scriptsize min}}\right)(\nu) 
&=&  \frac{1}{(2\pi)^\frac{3}{2}\,\theta^{*2}}
\exp\left[-\frac{\nu^2}{2}\right] \,\\
& &\nonumber  \times
\left[\left(\alpha^2(1-\gamma^2)-1+\gamma^2\nu^2\right)
+\frac{\exp\left[\frac{-\alpha^2\,\gamma^2\,\nu^2}{1+2\alpha^2(1-\gamma^2)}\right]}
{\sqrt{2\alpha^2(1-\gamma^2)+1}} \right]
\hspace{5pt}.
\end{eqnarray}

In addition to the formulae given in Ref.~\refcite{Bond_Efstathiou_1987}, 
we have specified here analytical expressions for the densities
$N_{\hbox{\scriptsize max}}(x,e)$, $N_{\hbox{\scriptsize max}}(x)$,
$N_{\hbox{\scriptsize max}}(e)$, $N_{\hbox{\scriptsize max}}(\varepsilon)$
and $N_{\hbox{\scriptsize max}}(E)$.
It should be pointed out that the formulae
in Ref.~\refcite{Bond_Efstathiou_1987}
are obtained by the leading term of the Laurent series of our formulae 
at $\alpha^2=1$.
As discussed in Section
\ref{Cosmological_Dependence_of_Gaussian_random_CMB_map}, 
the parameter $\alpha$ contains information about the underlying cosmology.

\subsection[The ellipticity in the CMB]{The ellipticity in the CMB}
\label{appendix_Ellipticity_of_the_CMB}

In the following the formulae for the moments of the ellipticity $E$,
the  eccentricity $\varepsilon$ and the elongation $e$ at local maxima
of an isotropic and homogeneous Gaussian random temperature field
on a 2-sphere are derived.
In the case of local minima the resulting formulae are also valid.
The elongation which is considered in this Appendix
results from a Taylor expansion at local extrema
and can be computed from the eigenvalues
of the Hessian (\ref{xe_lambda1_lambda2}).
For this reason the elongation $e$ contains only information
from local maxima or local minima.
The same is valid for the ellipticity $E$ and the eccentricity $\varepsilon$,
because they are related to the elongation
by Eqs.~(\ref{Eq:def_ellip}) and (\ref{Eq:def_excen}), respectively.
This is a difference to the definition of the ellipticity
in Section \ref{Elongation_HS_CS}
where the eigenvalues of the inertia tensor of the area
enclosed by the contour lines are used to calculate the ellipticity.

Using the mean differential densities of the maxima with respect to
the various arguments derived in \ref{appendix_peak_density},
one can define the distribution of a variable $u$ subject to
the constraint parameter $z$,
\begin{equation}
\label{Vert_P}
P(u|z) \; = \;
\frac{N_{\hbox{\scriptsize max}}(u,z)}{N_{\hbox{\scriptsize max}}(z)}
\hspace{10pt},
\end{equation}
e.g. $P(e|\nu)=N_{\hbox{\scriptsize  
max}}(\nu, e)/N_{\hbox{\scriptsize max}}(\nu)$ with the variable $e$  
and the parameter $\nu$.

The expectation values of the moments of the ellipticity with respect
to the distribution 
\begin{eqnarray} \nonumber
\label{P_e}
P(e)=
\frac{N_{\hbox{\scriptsize max}}(e)}{N_{\hbox{\scriptsize max}}} 
&=& \frac{24\, e (1-4e^2)\alpha^4 \sqrt{1+2\alpha^2}}{\left(1+8\alpha^2e^2\right)^{\frac{5}{2}}
\left[(\alpha^2-1)\sqrt{1+2\alpha^2}+1\right]} 
\\ &=& \hspace{0pt}
\frac{16\, e (1-4e^2)}{\left(1+8\alpha^2e^2\right)^{\frac{5}{2}}\phantom{i}_{2}F_1(1,\frac{5}{2};3;-2\alpha^2)}
\hspace{5pt}
\end{eqnarray}
are given by
\begin{equation}
\label{momente_e}
\langle e^n \rangle
= \int_{0}^{\frac{1}{2}}de\, P(e)\,e^n
=
\frac{B(\frac{n}{2}+1,2)\,_2F_1(\frac{n}{2}+1,\frac{5}{2};\frac{n}
{2}+3;-2\alpha^2)}{2^{n-1}\,_2F_1(1,\frac{5}{2};3;-2\alpha^2)}
\hspace{5pt}.
\end{equation}
$B(x,y)$ is the beta function.
We get $P(e)$ by using Eqs.~(\ref{N_max_e}) and (\ref{N_max})
and the moments by applying the integral~(2.2.6.15) in 
Ref.~\refcite{Prudnikov_Brychkov_Marichev_1988_Band1}.
From the expression of the moments, the mean value ($n=1$) results in
\begin{eqnarray}
\label{mittelwert_e}
\hspace*{-15pt}\langle e \rangle
& = &
\frac{2\alpha^2+3-\frac{3}{2}\sqrt{\frac{1+2\alpha^2}{2\alpha^2}}  \ln
\left[\frac{1+\sqrt{\frac{2\alpha^2}{1+2\alpha^2}}}{1-\sqrt{\frac{2\alpha^2}
{1+2\alpha^2}}}\right]}{4 \left[ (\alpha^2-1) \sqrt{1+2\alpha^2} +1
\right]} \\
\nonumber
& = & 0.197 - 0.041\,\,(\alpha ^{2}-1)+ \hbox{ O}((\alpha ^{2}-1)^2)
\hspace{10pt}
\end{eqnarray}
and the second moment ($n=2$) in
\begin{eqnarray}
\label{zweite_moment_e}
\langle e^2 \rangle
& = & \frac{1-3\,\sqrt{1+2\alpha^2} + 4
(\sqrt{1+2\alpha^2}-1)\left(\frac{1+2\alpha^2}{2\alpha^2}\right)}{4 \left[ (\alpha^2-1)
\sqrt{1+2\alpha^2} +1 \right]} \\
\nonumber
& = & 0.049 - 0.018\,\,(\alpha ^{2}-1)+ \hbox{ O}((\alpha ^{2}-1)^2)
\hspace{5pt}
\end{eqnarray}
of the ellipticity at maxima.

Using Eqs.~(\ref{N_max_epsilon}) and (\ref{N_max}) we obtain
the distribution
\begin{eqnarray}
\label{P_epsilon}
\nonumber
P(\varepsilon) &\;=\;&
\frac{N_{\hbox{\scriptsize max}}(\varepsilon)}{N_{\hbox{\scriptsize max}}} \\
&\;=\;& \frac{96\,\alpha^4 \sqrt{1+2\alpha^2}\,\varepsilon^3(1-\varepsilon^2)}
{\left[(\alpha^2-1)\sqrt{1+2\alpha^2}+1\right]
\left(4-4\varepsilon^2+(1+2\alpha^2)\varepsilon^4\right)^{\frac{5}{2}}} 
\hspace{5pt}
\end{eqnarray}
from which the mean value
\begin{eqnarray} \nonumber
\label{erwartungswert_epsilon}
\langle \varepsilon \rangle & = & \int_{0}^{1}d\varepsilon\, P(\varepsilon)\, \varepsilon \\  
& = & 0.715 - 0.059\,\,(\alpha ^{2}-1)+ \hbox{ O}((\alpha ^{2}-1)^2)
\hspace{5pt}
\end{eqnarray}
and the second moment
\begin{eqnarray} \nonumber
\label{zweite_moment_epsilon}
\langle \varepsilon^2 \rangle & = & \int_{0}^{1}d\varepsilon\, P(\varepsilon)\, \varepsilon^2 \\  
\nonumber
& = & \frac{2\left(\alpha^2+1-\sqrt{1+2\alpha^2}\right)}{(\alpha^2-1)\sqrt{1+2\alpha^2}+1} \\
& = & 0.536 - 0.083\,\,(\alpha ^{2}-1)+ \hbox{ O}((\alpha ^{2}-1)^2)
\hspace{5pt}
\end{eqnarray}
of the eccentricity can be calculated.

Using Eqs.~(\ref{N_max_gE}) and (\ref{N_max})
one gets the distribution
\begin{eqnarray}
\label{P_gE}
P(E)&=& 
\frac{N_{\hbox{\scriptsize max}}(E)}{N_{\hbox{\scriptsize max}}} 
\\&=& \nonumber
\frac{96\,\alpha^4 \sqrt{1+2\alpha^2}\,E^3(E^2-1)}
{\left[(\alpha^2-1)\sqrt{1+2\alpha^2}+1\right]
\left(1+2\alpha^2+2(1-2\alpha^2)E^2+(1+2\alpha^2)E^4\right)^{\frac{5}{2}}} 
\hspace{5pt},
\end{eqnarray}
which leads to the mean value
\begin{eqnarray} \nonumber
\label{erwartungswert_E}
\langle E \rangle & = & \int_{1}^{\infty}dE\, P(E)\, E \\  
\nonumber
& = & \frac{2}{\sqrt{3}}-\sqrt{3}K(\sqrt{2}i)+\frac{2}{\sqrt{3}}E(\sqrt{2}i)\\
\nonumber
& \phantom{=} &
+\left[-2+\frac{10}{\sqrt{3}\,3}
+\left(-\frac{1}{2\sqrt{3}}+3\right)K(\sqrt{2}i) 
+\left(\frac{5}{6\sqrt{3}}-2\right)E(\sqrt{2}i)
\right]\,(\alpha ^{2}-1)\\
\nonumber
& \phantom{=} & + \hbox{ O}((\alpha ^{2}-1)^2) \\
& = & 1.648 - 0.217\,\,(\alpha ^{2}-1)+ \hbox{ O}((\alpha ^{2}-1)^2)
\hspace{5pt}
\end{eqnarray}
and the second moment of the ellipticity $E$
\begin{eqnarray} \nonumber
\label{zweite_moment_E}
\langle E^2 \rangle & = & \int_{1}^{\infty}dE\, P(E)\, E^2 \\  
\nonumber
& = & \frac{(1+2\alpha^2)^{\frac{3}{2}}(\alpha^2+1)+4\alpha^4-4\alpha^2-1}{\left[(\alpha^2-1)\sqrt{1+2\alpha^2}+1\right](1+2\alpha^2)} \\
& = & 3.131 - 0.980\,\,(\alpha ^{2}-1)+ \hbox{ O}((\alpha ^{2}-1)^2)
\hspace{5pt}.
\end{eqnarray}
Here $K(k)$ is the complete elliptic integral of the first kind,
$E(k)$ the complete elliptic integral of the second kind and
$k$ the modulus.
 
All these moments of the ellipticity $E$, the elongation $e$ 
and the eccentricity $\varepsilon$ depend on $\alpha$. 
On the other hand $\alpha$ is determined by the angular power spectrum of the
underlying model, see Eq.~(\ref{Eq:alpha}).
For this reason also the moments of the ellipticity $E$, 
the elongation $e$ and 
the eccentricity $\varepsilon$
depend on the angular power spectrum of the model.
This dependence is considered in
Section \ref{Ellipticity_at_Maxima_and_Minima}.


\section*{Acknowledgements}

We would like to thank the Deutsche Forschungsgemeinschaft
for financial support (AU 169/1-1).
H.~S.~J.\ would like to thank the graduate school
``Analysis of complexitity, information and evolution''
of the Land Baden-W\"urttemberg for the stipend.
CMBFAST (www.cmbfast.org),
HEALPix\cite{Gorski_Hivon_Banday_Wandelt_Hansen_Reinecke_Bartelmann_2005}
(healpix.jpl.nasa.gov)
and the WMAP data from the LAMBDA website (lambda.gsfc.nasa.gov)
were used in this work.

\bibliography{../bib_astro}

\bibliographystyle{h-physrev5}

\end{document}